\documentclass[aps,prb,twocolumn,superscriptaddress]{revtex4-2} 

\usepackage[]{graphicx}
\usepackage[dvipsnames]{xcolor}
\usepackage[colorlinks=true,citecolor=NavyBlue,linkcolor=RubineRed,urlcolor=Cerulean,pdfencoding=auto]{hyperref}
\usepackage[]{amsmath}
\usepackage[]{amssymb}
\usepackage{multirow}
\def\bra#1{\mathinner{\langle{#1}|}}
\def\ket#1{\mathinner{|{#1}\rangle}}
\def\braket#1{\mathinner{\langle{#1}\rangle}}

\def\abs#1{\mathinner{|{#1}|}}

\def\proj#1{\ket{#1}\bra{#1}}

\def\comm#1#2{\mathinner{[{#1},{#2}]}}

\DeclareMathOperator{\I}{\openone}
\DeclareMathOperator{\atan}{arctan}
\DeclareMathOperator{\sign}{sign}

\begin{document}


\title{Strain Engineering for Transition Metal Defects in SiC}

\author{Benedikt Tissot}
\email[]{benedikt.tissot@uni-konstanz.de}
\affiliation{Department of Physics, University of Konstanz, D-78457 Konstanz, Germany}

\author{Péter Udvarhelyi}
\author{Adam Gali}
\affiliation{HUN-REN Wigner Research Centre for Physics, P.O.\ Box 49, H-1525 Budapest, Hungary}
\affiliation{Budapest University of Technology and Economics, Institute of Physics, Department of Atomic Physics, M\H{u}egyetem rakpart 3., 1111 Budapest, Hungary}

\author{Guido Burkard}
\email[]{guido.burkard@uni-konstanz.de}
\affiliation{Department of Physics, University of Konstanz, D-78457 Konstanz, Germany}


\begin{abstract}
  Transition metal (TM) defects in silicon carbide (SiC) are a promising platform for applications in quantum technology as some of these defects, e.g. vanadium (V), allow for optical emission in one of the telecom bands.
  For other defects it was shown that straining the crystal can lead to beneficial effects regarding the emission properties. Motivated by this, we theoretically study the main effects of strain on the electronic level structure and optical electric-dipole transitions of the V defect in SiC.
  In particular we show how strain can be used to engineer the \(g\)-tensor, electronic selection rules, and the hyperfine interaction.
  Based on these insights we discuss optical Lambda systems and a path forward to initializing the quantum state of strained TM defects in SiC.
\end{abstract}

\maketitle

\section{Introduction}

A fundamental ingredient for many quantum technologies and experiments is a coherent interface between flying qubits and stationary quantum memories~\cite{aharonovich16,heshami16,awschalom21}.
An established set of physical systems with great potential in this domain are so called color centers which are defects in solids with optical transitions.
Color centers can often additionally be coupled to nearby nuclear spins that lend themselves to quantum memories or long-lived quantum registers.

The most studied color center is the negatively charged nitrogen-vacancy (NV) defect in diamond \cite{he93,gaebel06,childress06,santori06,gali08,felton09,maze11,fuchs11,togan11,yale13,golter13,busaite20,hegde20}
(\cite{doherty13,suter17,pezzagna21} for reviews). 
Optical initialization and readout of its electron spin state is made feasible by the spin-photon interface via its excited state~\cite{Thiering_2018}. Together with a coherent microwave manipulation, optically detected magnetic resonance (ODMR) is feasible in this defect~\cite{Gruber_1997}. Efficient coupling to nearby nuclear spins was demonstrated and utilized in long living quantum memory applications \cite{fuchs11,busaite20,hegde20}. Despite its favorable spin and optical properties,   contenders for host materials other than diamond are emerging. The most notable is silicon carbide (SiC) with  advanced crystal growth~\cite{Wellmann_2018}, defect creation~\cite{Liu_2020, Chakravorty_2021}, and micro-fabrication techniques readily available~\cite{Song_2019, Lukin_2020, Guidry_2020}. These technological advancements improve the scalability~\cite{Radulaski_2017, Wang_2017} and magneto-optical properties of several hosted quantum defects, e.g. the negatively charged silicon vacancy~\cite{Sorman_2000, Janzen_2009, Nagy_2018} and the neutral divacancy~\cite{Gali_2010, Falk_2013}.

In this article, we focus on the transition metal (TM) defects in silicon-carbide (SiC), particularly on vanadium (V) defects.
In contrast to the defects discussed in the previous paragraph, V defects in SiC feature a zero-phonon line (ZPL) within the telecom bands, favorable for minimal loss transmission using optical fibers.
The focus of previous experiments~\cite{bosma18,spindlberger19,gilardoni20,wolfowicz20,astner22,cilibrizzi23} and theory~\cite{csore20,tissot21a,tissot21b,gilardoni21,tissot22} for TM defects in SiC was on unstrained defects, however, the knowledge on the external perturbations effecting the magneto-optical properties of the quantum defects is a key ingredient in their applications~\cite{udvarhelyi18, Udvarhelyi_2018, Udvarhelyi_2020, Udvarhelyi_2023}.
Strain  can be used passively, e.g. to reduce the dispersive readout time in silicon vacancy centers in diamond \cite{koppenhoefer23} 
and to engineer the electronic structure \cite{meesala18}
and $g$-tensor \cite{nguyen19},
or actively
to drive spin transitions in NV centers~\cite{udvarhelyi18}
as well as to create a hybrid quantum systems by coupling a mechanical oscillator to defects \cite{ovartchaiyapong14,barfuss19}.

Motivated by these prospects, in this work we aim to generalize the effective Hamiltonian to describe transition metal defects in silicon carbide under strain. 
To this end, we build on top of previous group-theory based results \cite{tissot21a,tissot21b}
which were in good agreement with previous \emph{ab-initio} calculations \cite{csore20}
and experimental findings \cite{wolfowicz20,gilardoni20,astner22,cilibrizzi23}.
Additionally, we use density functional theory (DFT) calculations to estimate the strain coupling strength for the commonly used vanadium defect in the $k$ site of 4H-SiC \footnote{
According to previous DFT results the $k$ site corresponds to the \(\alpha\) configuration of vanadium in 4H-SiC \cite{csore20}.
}.
We show how strain in these samples can be used to engineer the optical transition frequency,
the $g$-tensor, transition rules as well as the form of the hyperfine interaction.
Based on this, we discuss state preparation and readout as well as microwave control in strained samples.

This paper is organized as follows. We begin by introducing the physical model for the V defect in SiC in Sec.~\ref{sec:model}, including its effective Hamiltonian. 
Using this model, we combine and compare the effective Hamiltonian and \emph{ab initio} calculations in Sec.~\ref{sec:abinit}.
Based on these results, we then show the possibility to engineer the $g$-tensor (Sec.~\ref{sec:gtensor}), selection rules (Sec.~\ref{sec:trans}), and how these can be combined to create a Lambda system for pseudo-spin state preparation (Sec.~\ref{sec:pol}).
In Sec.~\ref{sec:hyperfine}, we discuss the influence of strain on the hyperfine interaction and how this influences the possibility to initialize the nuclear spin. 
We summarize our findings and present our conclusions in Sec.~\ref{sec:conclusion}.

\section{Model\label{sec:model}}

\subsection{Defect structure}

\begin{figure}[t]
\centering
\includegraphics[width=\linewidth]{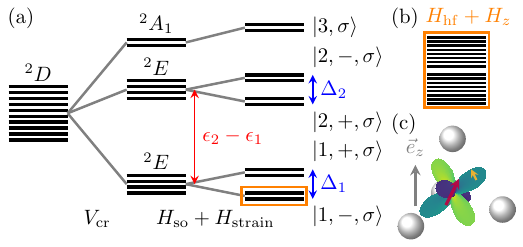}
\caption{
\label{fig:sketch}
Sketch of the level structure and atomic configuration at the defect site.  (a) Hierarchy leading to the strained electronic structure, where the largest splitting 
of the $D$-shell levels occupied by a single electron
is due to the crystal field (red arrow), leading to two orbital doublets $E$ and one orbital singlet $A_1$, with additional two-fold spin degeneracy.
The doublets are further split due to the combination of strain and spin-orbit interaction, resulting in the leading-order splitting 
[see Eq.~\eqref{eq:Edoublet}]
between the Kramers doublets (KDs).
The KD originating from the orbital singlet \(A_1\) is not further split.
Zooming into one of the KDs, (b) reveals the hyperfine structure and Zeeeman splitting.
(c) Artistic illustration of the \(D\) shell electronic orbital of the defect (green and purple), electron spin (yellow), nuclear spin (dark red), and nearest neighboring sites (gray balls). The gray arrow indicates the crystal axis \(\vec{e}_z\).}
\end{figure}

The defect energy levels, sketched in Fig.~\ref{fig:sketch}, can be described by a single electron in an orbital resembling the original atomic $d$  orbital.
The \(^2D\) levels are split by the crystal potential into two orbital doublets $^2E$ and one orbital singlet $^2A_1$.
Due to the spin-orbit interaction and the interaction with an external strain field, the orbital doublets are further split.
This results in a level structure made up by five Kramers doublets (KDs) which are pairs of states related to each other by time inversion.
We use a group theoretic model in the following to describe the above interactions within an effective Hamiltonian where we additionally calculate selection rules between the KDs, the hyperfine structure of the KDs, and the Zeeman term within each KD.
Therein, we calculate the orbital-strain interaction parameters not yet reported in the literature using \emph{ab initio} calculations and also use these to confirm predictions made by the effective Hamiltonian.

\subsection{Effective Hamiltonian\label{sec:Heff}}
We generalize the effective Hamiltonian 
from \cite{tissot21a,tissot21b,tissot22}
to additionally include the strain such that the full Hamiltonian is of the form
\begin{align}
  \label{eq:Htotal}
  H = H_{\mathrm{TM}} + V_{\mathrm{cr}} + {H}_{\mathrm{so}} + {H}_{z} + H_{\mathrm{nuc}} + H_{\mathrm{el}} + H_{\mathrm{st}},
\end{align}
where the form of the atomic Hamiltonian (\(H_{\mathrm{TM}}\)), crystal potential \(V_{\mathrm{cr}}\), spin-orbit interaction \(H_{\mathrm{so}}\), and coupling to magnetic \(H_z\) and electric fields \(H_{\mathrm{el}}\) are discussed in \cite{tissot21a}.
The interaction with the central nuclear spin \(H_{\mathrm{nuc}}\) (of the TM) was derived and analyzed in \cite{tissot21b}.
We summarize the form of these terms below and additionally discuss the influence of strain \(H_{\mathrm{st}}\) within this symmetry-based framework.
We note that this approach can only work in the domain where the strain can be viewed as a small perturbation compared to \(H_{\mathrm{TM}} + V_{\mathrm{cr}}\). As we show in the following, this restriction does not affect our conclusions, since significant strain effects are demonstrated within this domain.

In the basis of (orbital) eigenstates \(\ket{0}, \ket{\pm_{n}}\) with \(n=1,2\) of \(H_{\mathrm{TM}} + V_{\mathrm{cr}}\) (introduced in \cite{tissot21a}) and using the corresponding projections of degenerate subspaces
\(P_n = \proj{+_n} + \proj{-_n}\) and \(P_3 = \proj{0}\),
we can write the combined atomic and crystal Hamiltonian as
\begin{align}
  \label{eq:Hcrystal}
   H_{\mathrm{TM}} + V_{\mathrm{cr}} = \sum_{n=1,2,3} \epsilon_n P_n,
\end{align}
where \(\epsilon_n\) are the crystal energies.
Here, $n=1,2,3$ labels the three orbital multiplets shown in Fig.~\ref{fig:sketch}(a).
In the following, we describe each part of the Hamiltonian $H$ by its contribution to each of the nine blocks $P_n H P_m$ defined by the three orbital sectors ($n,m=1,2,3$).

Using the projection operators, we can formulate the different blocks of the spin-orbit interaction as
\begin{align}
  \label{eq:Hso}
  P_n H_{\mathrm{so}} P_m = &\, \lambda_{nm}^z S_z \sigma_z + i (n-m) \lambda_{nm}^x (S_x \sigma_y + S_y \sigma_x) , \\
  \braket{ 0 | H_{\mathrm{so}} | \pm_n } = &\, \lambda_{n3}^x (S_x \pm i S_y), \quad \braket{ 0 | H_{\mathrm{so}} | 0 }  = 0
\end{align}
where \(n,m = 1,2\), \(\sigma_{k}\) denote the Pauli matrices (here acting between the orbital states states $\ket{\pm_n}$), and \(S_k\) (\(k=x,y,z\)) the spin-1/2 operators of the electron in units of \(\hbar\).
The spin-orbit coupling constants \(\lambda_{nm}^k\) are in units of energy
with this convention.

Another relevant term is the electronic Zeeman interaction due to the spin and angular momentum coupling to an external magnetic field.
In the relevant subspaces, it is given by
\begin{align}
  \label{eq:Hz}
  P_n H_{z} P_m = &\, r_{nm}^z B_z \sigma_z +  \mu_B g_s \vec{B} \cdot \vec{S} \delta_{nm} \notag \\
    & + i (n-m) r_{nm}^x (B_x \sigma_y + B_y \sigma_x), \\
  \bra{0} H_{z} \ket{\pm_n} = &\, r_{n3}^x (B_x \pm i B_y), \, \bra{0} H_{z} \ket{0} = \mu_B g_s \vec{B} \cdot \vec{S} ,
\end{align}
where we use the electron gyromagnetic ratio \(g_s = 2\), the Bohr magneton \(\mu_B\), and the 
coupling constants of the orbital Zeeman term \(r_{nm}^k\).

The nuclear Hamiltonian \(H_{\mathrm{nuc}} = H_{\mathrm{hf}} + H_{\mathrm{zn}}\) (for transition metal defects with non-zero nuclear spin)
is made up by the hyperfine interaction \(H_{\mathrm{hf}}\) and the nuclear Zeeman interaction \(H_{zn} = \mu_N g_N \vec{B} \cdot \vec{I}\) with the nuclear spin operator \(\vec{I}\) in units of \(\hbar\), the nuclear magneton \(\mu_N\), and nuclear \(g\)-factor \(g_N\).
As the nuclear Zeeman term is proportional to the identity operator \(\I\) in the electronic subspace, it can be straightforwardly incorporated.
Following \cite{tissot21b} and accommodating to the notation employed in this article, the hyperfine interaction projected onto the relevant subspaces is
\begin{widetext}
\begin{align}
  P_n H_{\mathrm{hf}} P_m = &\,
 \left[  a_{nm}^z S_z I_z + \frac{{a_{nm}^z}'}{2} \left(S_+ I_- + S_- I_+\right) \right]
    + \left\{ - \frac{a_{nm}^x}{2} \left[ \sigma_- S_+ I_+ + \sigma_+ S_- I_- \right] \right. \notag \\
  \label{eq:Hhf}
  & \left. + \frac{1}{2} {a_{nm}^x}' \left[ S_z \left(\sigma_+ I_+ + \sigma_- I_-\right) + \left( S_+ \sigma_+ + \sigma_-  S_- \right) I_z \right] \right\}
    + {a_{nm}^z}'' I_z \sigma_z + \frac{{a_{nm}^x}''}{2} (n-m)  \left(\sigma_+ I_+ - \sigma_- I_-\right), \\
  \label{eq:Hhf2}
  \bra{0} H_{\mathrm{hf}} \ket{\pm_n} = &\, {a_{n3}^x}'' I_{\pm}  \pm a_{n3}^x S_{\mp} I_{\mp} \mp {a_{n3}^x}' \left[ S_z I_{\pm} + S_{\pm} I_z \right] ,
   \qquad \bra{0} H_{\mathrm{hf}} \ket{0} = a_{33}^z S_z I_z + \frac{1}{2} {a_{33}^z}' \left(S_+ I_- + S_- I_+\right) ,
\end{align}
\end{widetext}
where we use the ladder operators \(O_{\pm} = O_x \pm i O_y\) with \(O=\sigma,S,I\)
as well as the set of 
hyperfine coupling parameters
\(a_{nm}^{k}\), \({a_{nm}^{k}}'\), and \({a_{nm}^{k}}''\).

The effect of an applied electric field can be described by
\begin{align}
  \label{eq:Hel}
  P_n H_{\mathrm{el}} P_m = &\, \mathcal{E}_{nm}^z E_z \I + \mathcal{E}_{nm}^z ( \sigma_x E_x - \sigma_y E_y ) , \\
  \label{eq:Hel2}
  \bra{0} H_{\mathrm{el}} \ket{\pm_n} = &\, \mp \mathcal{E}_{n3}^x (E_x \pm i E_y) , \, \bra{0} H_{\mathrm{el}} \ket{0} = \mathcal{E}_{33}^z E_z,
\end{align}
with the 
coupling strengths \(\mathcal{E}_{nm}^k\).

Lastly, we turn to the strain Hamiltonian.
In our model, we start with a product space of orbital and spin components, such that we can incorporate the strain interaction within the orbital subspace.
Its competition with the spin-orbit interaction gives rise to a complex interplay within the KDs.
To describe it, we use the assignment of the different strain elements to irreducible representations of \(C_{3v}\) \cite{udvarhelyi18} which then couple to the corresponding orbital operator, \(i.e.\) the strain components transforming like the basis \(\{x,y\}\) of the irreducible representation \(E\) couple to operators of the form of \(x\) and \(y\), while the strain components transforming like the basis \(z\) of \(A_1\) couple to operators of the form of \(z\).
With these considerations, the strain Hamiltonian
\begin{align}
  \label{eq:Hstrain}
  P_n H_{\mathrm{st}} P_m =\, & \epsilon_{nm}^z \I + ( \sigma_x \epsilon_{nm}^x - \sigma_y \epsilon_{nm}^y ) , \\
  \label{eq:Hstrain2}
  \bra{0} H_{\mathrm{st}} \ket{\pm_n} =\, & \mp (\epsilon_{n3}^x \pm i \epsilon_{n3}^y) , \quad \bra{0} H_{\mathrm{st}} \ket{0} = \epsilon_{33}^z,
\end{align}
has a similar structure as the coupling to electric fields but potentially leads to a much larger contribution.
Here, we use the reduced components of the strain tensor, organized by symmetry, \(\epsilon_{nm}^x = s_{nm}^{x} \epsilon_{xz} + {s_{nm}^{x}}' \frac{\epsilon_{yy} - \epsilon_{xx}}{2}\), \(\epsilon_{nm}^y = s_{nm}^{x} \epsilon_{yz} + {s_{nm}^{x}}' \epsilon_{xy}\), and \(\epsilon_{nm}^z = s_{nm}^{z} \epsilon_{zz} + {s_{nm}^{z}}' \frac{\epsilon_{xx} + \epsilon_{yy}}{2}\).
In contrast to the coupling to electric (and magnetic) fields, the tensorial form of the strain manifests itself in the presence of multiple strain elements pertaining to the same irreducible representation.
These elements can also have different coupling constants \(s_{nm}^k\) and  \({s_{nm}^k}'\) leading to more degrees of freedom than a coupling to vectors.

For concreteness, we focus on the vanadium defect in the \(k\) site of 4H-SiC in the following. We use the already known combinations of parameters 
and we additionally estimate the magnitude of the strain coupling constants using DFT calculations.
Many relevant 
parameters without strain can be found
in our previous works \cite{tissot21a,tissot21b,tissot22},
including their values for other defects.

\section{Results\label{sec:results}}
In a first step, we investigate the electronic structure following from an externally applied, uniaxial, and static strain.
The remaining terms in the Hamiltonian will be discussed afterwards, omitting the discussion of static electric fields as they couple weakly to the defect compared to strain and magnetic fields and the symmetry-based  electric-field coupling Hamiltonian [Eqs.~\eqref{eq:Hel}~and~\eqref{eq:Hel2}] is similar to the strain Hamiltonian [Eqs.~\eqref{eq:Hstrain}~and~\eqref{eq:Hstrain2}].

Referring to the absence of an external electromagnetic field as ``zero field'', we define the electronic zero-field Hamiltonian as
\(H_{\mathrm{ezf}} = H_{\mathrm{TM}} + V_{\mathrm{cr}} + H_{\mathrm{so}} + H_{\mathrm{st}}\).
Projected onto one of the doublets, the electronic zero-field Hamiltonian is
\begin{align}
  \label{eq:Hezf}
  P_n H_{\mathrm{ezf}} P_n = & (\epsilon_n + \epsilon_{nn}^z) \I + \lambda_{nn}^z S_z \sigma_z
  + (\sigma_x \epsilon_{nn}^x - \sigma_y \epsilon_{nn}^y) ,
\end{align}
where we take the crystal field splitting to be the dominant contribution, i.e. \(\abs{\epsilon_n - \epsilon_m} \gg \lambda_{jl}^k, \epsilon_{jl}^k\) with \(n \ne m\), \(n,m,j,l=1,2,3\), and \(k=x,y,z\).
Furthermore, we investigate the domain where the magnetic field is weak compared to the spin-orbit coupling, as is relevant for most experimental and technological applications.
Therefore, we begin by diagonalizing the Hamiltonian~\eqref{eq:Hezf}, leading to the eigenvalues
\begin{align}
  \label{eq:Edoublet}
  E_{n,\pm} = \epsilon_n + \epsilon_{nn}^z \pm \frac{1}{2} \sign(\lambda_{nn}^z) \Delta_n,
\end{align}
with the combined spin-orbit and strain splitting ${\Delta_n = \sqrt{(\lambda_{nn}^z)^2 + 4(\epsilon_{nn}^x)^2+ 4(\epsilon_{nn}^y)^2}}$.
As it splits the orbital doublet into two Kramers doublets (made up by two pseudo spins) we refer to $\Delta_n$ as the orbital splitting.
These energies are doubly degenerate in agreement with Kramers' theorem, as time-reversal symmetry is still preserved for this static Hamiltonian, despite the (potential) spatial symmetry breaking due to strain.
The corresponding eigenstates are
\begin{align}
  \ket{n,\pm,\sigma} = &\, \cos(\theta_n/2) \ket{{\pm\sigma}_n} \ket{\sigma} \notag \\
  \label{eq:StrainStatesDoublet}
  &\pm \sin(\theta_n/2) \exp(\mp i \sigma \varphi_n/2) \ket{{\mp\sigma}_n} \ket{\sigma} ,
\end{align}
where \(\sigma = \uparrow,\downarrow\) denotes the spin and is also used as \(\sigma = \pm\) to achieve a concise notation.
The \(x\)- and \(y\)-like components of the strain coupling compete with each other and with the spin-orbit coupling,
leading to the mixing angles \(\tan(\theta_n) = 2\epsilon_{nn}^x \sqrt{1 + (\epsilon_{nn}^y/\epsilon_{nn}^x)^2} \Big/ \lambda_{nn}^z\) and
\(\tan(\varphi_n) = \epsilon_{nn}^y/\epsilon_{nn}^x\).
Without strain, the \(C_{3v}\) symmetry of the defect is intact, yielding the KDs \(\Gamma_4\) and \(\Gamma_{5/6}\) for the \(\ket{n,-,\sigma}\) and \(\ket{n,+,\sigma}\) KDs, respectively.

\subsection{\emph{Ab initio} calculations\label{sec:abinit}}

The defect structure shows $\text{C}_{3\text{v}}$ point symmetry owing to the axial crystal field of the 4H polytype.
It introduces a double-degenerate $e^{(1)}$
orbital inside the band gap, occupied by a single electron and two empty orbital levels ($e^{(2)}$ and $a_1$) which are localized inside the conduction band in the $^{2}E$ ground state electronic configuration. The lowest energy excitation promotes the electron between the different $e$ levels, sinking the $e^{(2)}$ orbital inside the band gap. The calculated ZPL energy of 0.91 eV is in reasonable agreement with experiments~\cite{wolfowicz20}. We note that Jahn-Teller instabilities are suppressed in the calculations by a smeared occupation in the $e$ orbital subspace, describing a dynamically averaged system in the unperturbed solution and strain perturbation is applied to this high-symmetry system.

First, we determine the $s_{nm}^k$ orbital-strain coupling constants, without spin-orbit coupling taken into account, in both the ground and first excited state of the defect. To this end, we apply strain with a magnitude of up to 0.02 and fit a linear response for the orbital level splitting energy and the ZPL energy in the case of $E$ and $A_1$ strain components, respectively.
The coupling coefficients are extracted as the fitted slope.
Within the margin of error (see App.~\ref{app:DFT}) the slope for strains transforming together agree and are predicted to have the same $s_{nm}^k$ by the effective Hamiltonian,
such that we will use their average in the following.
The calculated values are collected in Table~\ref{tab:strain_coupling}
where we additionally assigned the signs based on the discussion in App.~\ref{app:strainsign}.

\begin{table}[]
    \centering
    \begin{tabular}{l r|r|r}
      $n$ & $k$ & $s_{nn}^k$  (${h\,\mathrm{THz}}/{\mathrm{strain}}$) & ${s_{nn}^k}'$ (${h\,\mathrm{THz}}/{\mathrm{strain}}$) \\\hline 
      $1$ & $x$ & $251 \pm 1$ & $230 \pm 3$ \\
      $2$ & $x$ & $-138 \pm 6$  & $-204 \pm 3$ \\
      $2$ & $z$ & $459 \pm 24$ & $305 \pm 19$ \\
    \end{tabular}
    \caption{
    \label{tab:strain_coupling}
    Calculated strain-orbital coupling coefficients extracted from the linear perturbation model of the orbital level splitting and ZPL energies from DFT.
    We average the coupling of elements pertaining to the same irreducible representation as they are predicted to be the same by the Wigner-Eckart theorem and agree within numerical accuracy between the original DFT results, see Tab.~\ref{tab:DFT_strain}.
    The signs are assigned according to App.~\ref{app:strainsign}.
    For the effective Hamiltonian we choose to use $s_{11}^z = {s_{11}^z}'=0$ and assign the 
    slope of the energy difference from the DFT calculations to $s_{22}^z$ and ${s_{22}^z}'$.
    }
\end{table}

\begin{figure}[t]
\centering
\includegraphics[width=\linewidth]{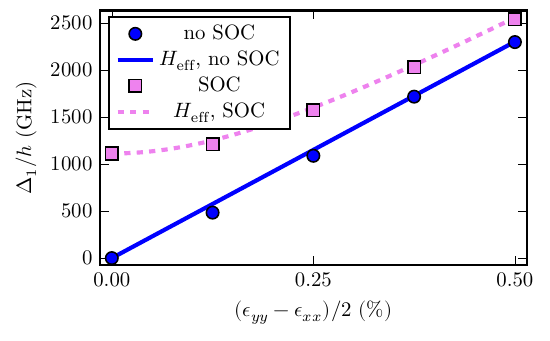}
\caption{\label{fig:DFT}
Ground-state orbital splitting $\Delta_1$ as a function of strain $\epsilon_{yy} - \epsilon_{xx}$ comparing the effective Hamiltonian and DFT calculations.
The blue line (dots) shows the linear dependence of the orbital splitting in the absence of spin-orbit splitting obtained from the effective Hamiltonian (from DFT).
The pink dashed line (squares) show the combined spin-orbit and strain splitting from the effective Hamiltonian (from DFT), following the prediction, Eq.~\eqref{eq:Edoublet}.
The effective Hamiltonian uses the zero-strain spin-orbit splitting (extracted from the DFT) and fitted ${s_{11}^x}'$ (see Tab.~\ref{tab:strain_coupling}).
}
\end{figure}
To confirm the agreement between the effective Hamiltonian and the \emph{ab initio} calculations as well as the extrapolation of using the purely orbital calculations to extract the strain coupling constants $s^k_{nm}$ we use these combined with the spin-orbit splitting in the absence of strain predicted by the \emph{ab initio} calculation given by about $\lambda_{11}^z/h = 1112\,$GHz to calculate the ground state energy splitting and compare them to simulations combining strain and the spin-orbit interaction, see Fig.~\ref{fig:DFT}.
The spin-orbit splitting calculated using DFT turns out to be about twice
the experimentally measured value of $529\,$GHz~\cite{wolfowicz20} in agreement with a Ham reduction factor of about $0.6$~\cite{csore20}.

\begin{figure}[t]
\centering
\includegraphics[width=\linewidth]{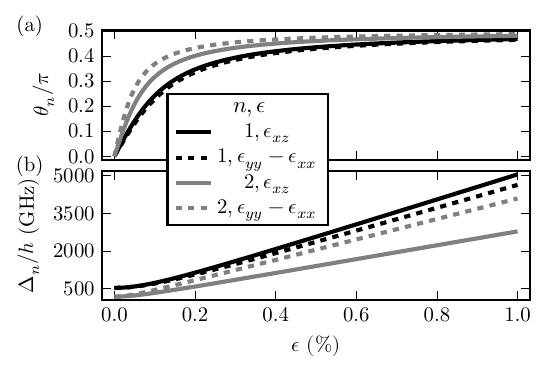}
\caption{
\label{fig:splitting}
Behavior of Kramers doublets pertaining to the same orbital doublet under strain $\epsilon$ at zero field.
(a) Strain mixing angle \(\theta_n\) [see Eq.~\eqref{eq:StrainStatesDoublet}], and (b) orbital energy splitting [see Eq.~\eqref{eq:Edoublet}] as a function of $\epsilon$.  
The plots show an increase in both mixing and splitting with increasing strain. The symmetric KDs $\Gamma_4$ and $\Gamma_{5/6}$ are fully mixed for highly strained samples.
Furthermore, we note that the mixing angle is antisymmetric, while the energy splitting is symmetric when inverting the sign of the strain.
We use the coupling constants in Tab.~\ref{tab:strain_coupling} as well as 
$\lambda_{11}^z/h = 529\,$GHz and $\lambda_{22}^z/h = -181\,$GHz.
}
\end{figure}

For this reason, we will use the numerically determined strain coupling constants but use experimentally determined parameters from the literature where available.
In particular, we compare the mixing angle as well as the combined spin-orbit and strain splitting as functions of strain components transforming according to $E_x$ of the GS and ES doublets in Fig.~\ref{fig:splitting} using the experimentally determined spin-orbit splittings $\lambda_{11}^z/h = 529\,$GHz and $\lambda_{22}^z/h = -181\,$GHz~\cite{wolfowicz20}, where we assigned the signs based on the level ordering \cite{tissot21b}.
Fig.~\ref{fig:splitting} shows that the splittings of the ES and GS diverge more for $\epsilon_{yy} - \epsilon_{xx}$ strain than for $\epsilon_{xz}$.
The mixing angle between the strain types is also different where it increases faster for $\epsilon_{xz}$ strain and in all cases approach the asymptotic value of $\pi/2$ which corresponds to maximal mixing of the unstrained KDs, see~Eq.~\eqref{eq:StrainStatesDoublet}.

The linear dependence of Eq.~\eqref{eq:Edoublet} on $z$-type strain, i.e.~$\epsilon_{zz}$ and $\epsilon_{xx}+\epsilon_{yy}$, 
combined with the non-zero difference of the coupling constants between the GS and ES from the \emph{ab initio} calculation (see Tab.~\ref{tab:strain_coupling})
implies that $z$-type strain can be used to tune the optical transition frequency, i.e.~the crystal field splitting.
This is possible while keeping the selection rules intact as $z$-type strain conserves the defect's $C_{3v}$ symmetry.
Within the effective Hamiltonian we choose to use $s_{11}^z = {s_{11}^z}'=0$ 
because the overall energy shift can be set arbitrarily and only the energy differences contained in the effective Hamiltonian carry physical meaning.
We assign the full difference of the coupling of the ES and GS orbital doublets extracted from the \emph{ab initio} calculation to the parameters $s_{22}^z$ and ${s_{22}^z}'$.

\subsection{Engineering the \(g\)-tensor\label{sec:gtensor}}

\begin{figure}[t]
\centering
\includegraphics[width=\linewidth]{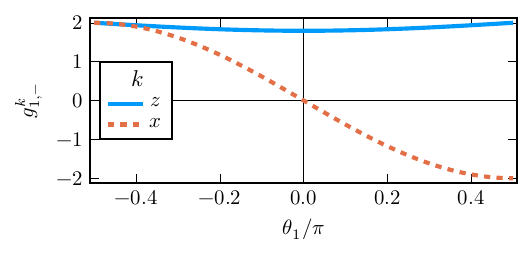}
\caption{\label{fig:gtensor} Parallel and perpendicular \(g\)-tensor elements as a function of the strain mixing angle \(\theta_1\) for the GS KD.
In the absence of strain the perpendicular $g$-factor vanishes and increases (decreases) for positive (negative) strain mixing angles $\theta_1$.
The parallel $g$-factor only slightly varies around the value of $2$.
Parameters used for this plot are $g_s = 2$ and $r_{11}^z = 0.103$.
}
\end{figure}

Projecting onto the \(n(=1,2),\pm\) KD (spanned by the states \(\ket{n,\pm,\sigma}\) with \(\sigma=\uparrow,\downarrow\)) we calculate the leading-order Zeeman term
\begin{align}
\label{eq:HzEff}
   & H_{n,\pm}^z =  \mu_B g_{n,\pm}^z \tilde{S}_z B_z \\
   \notag
   & \quad + \mu_B g_{n,\pm}^x \begin{pmatrix} \tilde{S}_x \\ \tilde{S}_y \end{pmatrix}
   \begin{pmatrix} \cos(\varphi_n) & \pm \sin(\varphi_n) \\ \mp\sin(\varphi_n) & \cos(\varphi_n) \end{pmatrix}
   \begin{pmatrix} B_x \\ B_y \end{pmatrix} ,
\end{align}
with the effective \(g\)-factors \(g_{n,\pm}^z = g_s \pm 2 r_{nn}^z \cos(\theta_n)\) and \(g_{n,\pm}^{x} = \pm g_s \sin(\theta_n)\),
where \(\tilde{S}_k\) is the \(k=x,y,z\) pseudo-spin (\(1/2\)) operator for the KD.
From this expression, it is evident that using \(x,y\) strain enables coupling to perpendicular magnetic fields such that quantum gates relying on \(S_{x,y}\) operators become possible using microwave drives.
Furthermore, \(y\)-type strain leads to an effective rotation of the spin in the \(x,y\) plane regarding an external magnetic field.

As we previously showed in the absence of strain \cite{tissot21a},
in the presence of strain
the \(g\)-factors are also influenced by (combined) higher orders of the strain and spin-orbit interactions between different orbital subspaces.
We calculate the (second-order) correction using a Schrieffer-Wolf transformation
treating \(H_{\mathrm{so}} + H_{\mathrm{st}}\) as the perturbation.
In App.~\ref{app:SW}, we show how to derive the correction for \(g_{n,\pm}^z\) for purely \(x\)-type strain. 
These are in agreement with previous unstrained results.
Using the insights of the higher order, we can calculate \(r_{nn}^z\) as the mean deviation from \(\sum_{s=\pm}| g_s - g_{n,s}^z|/2\) of the experimentally determined \(g_{n,s}^z\) of the same doublet \(n\) and attribute the remaining deviation to a common deviation from \(g_s\) due to the second-order term.
With this consideration and using the $g$-factors~\cite{tissot21b,wolfowicz20,astner22}, we find \(r_{11}^z = 0.103\) for vanadium defects in the \(k\) site in 4H-SiC (and the second order correction is \(0.046\)).
Using these parameters, we show the evolution of the parallel and perpendicular \(g\)-factors as a function of the strain mixing angle \(\theta_1\) for the ground-state KD in Fig.~\ref{fig:gtensor} where we do not include the small second-order correction.
This figure makes it readily visible that as \(r_{11}^z \ll g_s\) the parallel \(g\)-factor changes little compared to the perpendicular \(g\)-factor.
The perpendicular \(g\)-factor $g_{1,-}^x$ varies between \(0\) in the absence of strain up to \(\pm g_s =\pm 2\) 
while the parallel $g$-factor $g_{1,-}^z$ only shows a marginal deviation from $g_s = 2$.

\subsection{Engineering optical transitions\label{sec:trans}}

\begin{figure}[t]
\centering
\includegraphics[width=\linewidth]{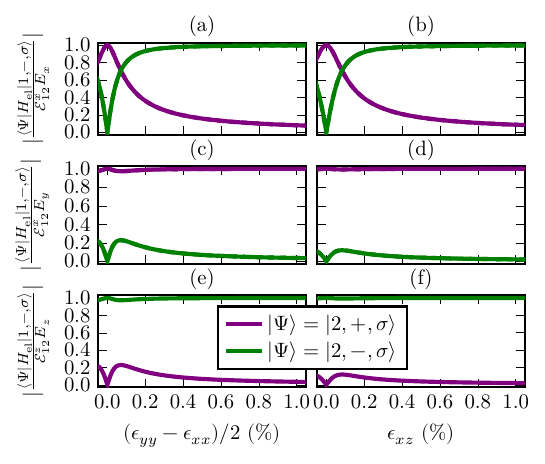}
\caption{\label{fig:dipoleTrans} Electric dipole transition matrix element as a function of \(x\)-type strain for a $z$ polarized drive.
The color (see legend) encodes the target ES $\ket{1,\pm,\sigma}$ that couples pseudo-spin conserving to the GS KD $\ket{1,-,\sigma}$.
The rows (a,b), (c,d), and (e,f) show the selection rules for $E_x, E_y, E_z$, respectively.
The columns (a,c,e) and (b,d,f) correspond to the two different $x$-type strains $(\epsilon_{yy}-\epsilon_{xx})/2$ and $\epsilon_{xz}$.
The pseudo-spin $\sigma$ is encoded in the line style solid (dashed) for $\uparrow$ ($\downarrow$) but is not visible as the lines are aligned.
We use the same parameters as in Fig.~\ref{fig:splitting}.
}
\end{figure}

After discussing the interaction with magnetic fields which is essential to split the pseudo-spin levels and for microwave control, 
we investigating the leading-order electric dipole transition matrix elements in this subsection.
These matrix elements are important to characterize the interaction with optical fields.
In the absence of strain
all leading-order transitions conserve the spin \(\braket{ n, \pm, \sigma | H_{\mathrm{el}} | m, \pm, -\sigma } = 0\)  and \(\braket{ n, \pm, \sigma | H_{\mathrm{el}} | 3, -\sigma } = 0\) \cite{tissot21a}.
For simplicity, we focus on the transitions between the \(1,-\) KD and the two \(2,\pm\) KDs under the (leading-order) influence of strain.
To this end, we calculate
\begin{widetext}
\begin{align}
  \label{eq:HelStrain1}
  & \braket{ 2, +, \sigma | H_{\mathrm{el}} | 1, -, \sigma } \sigma e^{-\sigma i (\varphi_1 + \varphi_2)/2} =
    E_z \mathcal{E}_{12}^z \left[ e^{i (\varphi_2 - \varphi_1)/2} \cos\left(\frac{\theta_1}{2}\right) \sin\left(\frac{\theta_2}{2}\right) - e^{- i (\varphi_2 - \varphi_1)/2} \cos\left(\frac{\theta_2}{2}\right) \sin\left(\frac{\theta_1}{2}\right) \right] \notag \\
  & \qquad + \mathcal{E}_{12}^x \left[ (\sigma E_x + i E_y) e^{-\sigma i (\varphi_2 + \varphi_1)/2} \cos\left(\frac{\theta_1}{2}\right) \cos\left(\frac{\theta_2}{2}\right) + (-\sigma E_x + i E_y) e^{\sigma i (\varphi_2 + \varphi_1)/2} \sin\left(\frac{\theta_1}{2}\right) \sin\left(\frac{\theta_2}{2}\right)  \right], \\
  \label{eq:HelStrain2}
  & \braket{ 2, -, \sigma | H_{\mathrm{el}} | 1, -, \sigma } e^{\sigma i (\varphi_2 - \varphi_1)/2} =
  E_z \mathcal{E}_{12}^z \left[ e^{\sigma i (\varphi_2 - \varphi_1)/2} \cos\left(\frac{\theta_1}{2}\right) \cos\left(\frac{\theta_2}{2}\right) + e^{- \sigma i (\varphi_2 - \varphi_1)/2} \sin\left(\frac{\theta_2}{2}\right) \sin\left(\frac{\theta_1}{2}\right) \right] \notag \\
  & \qquad + \mathcal{E}_{12}^x \left[ (- E_x - \sigma i E_y) e^{-\sigma i (\varphi_2 + \varphi_1)/2} \cos\left(\frac{\theta_1}{2}\right) \sin\left(\frac{\theta_2}{2}\right) + (- E_x + \sigma i E_y) e^{\sigma i (\varphi_2 + \varphi_1)/2} \sin\left(\frac{\theta_1}{2}\right) \cos\left(\frac{\theta_2}{2}\right)  \right] .
\end{align}
\end{widetext}
Similar expressions can be analogously calculated for other transitions.
We show in Fig.~\ref{fig:dipoleTrans} how the electric dipole transition matrix elements evolve as a function of the \(x\)-type strain elements \(\epsilon_{xz}\) and \((\epsilon_{yy}-\epsilon_{xx})/2\).
This figure underlines that there are domains where these types of strain enable multiple simultaneous transitions and that strain can significantly impact the selection rules of the defect.
The strongest change of the selection rules visible in Fig.~\ref{fig:dipoleTrans} is the complete inversion of dipole coupling strength to the GS via $E_x$ polarized fields between the ES KDs under $x$ strain.
Combined with the influence on $E_y$ we conclude that the circular polarization selection rules in the absence of strain~\cite{tissot22} become linear polarization rules in suitably strained samples.
For example, in Fig.~\ref{fig:dipoleTrans} it can be seen that the transition $\ket{1,-,\sigma} \leftrightarrow \ket{2,+,\sigma}$ becomes primarily susceptible to $E_y$ in the presence of strong strain.
We can generalize this by considering that for strong (positive) strain  $\theta_1,\theta_2 \approx \pi/2$,
such that we find 
\begin{widetext}
\begin{align}
\label{eq:melHel_strong_strain}
& \braket{ 2, +, \sigma | H_{\mathrm{el}} | 1, -, \sigma } \sigma e^{-\sigma i (\varphi_1 + \varphi_2)/2} 
= 
i E_z \mathcal{E}_{12}^z \sin \left( \frac{\varphi_2 - \varphi_1}{2} \right) - i \mathcal{E}_{12}^x \left[ E_x \sin \left( \frac{\varphi_2 + \varphi_1}{2} \right) - E_y \cos \left( \frac{\varphi_2 + \varphi_1}{2} \right)  \right].
\end{align}
\end{widetext}
Figure~\ref{fig:dipoleTrans} and the above expressions directly show that we can generate an orbital three-level system in the \(V\) configuration where one GS KD couples to two ES KDs in the presence of strain.
Considering the equivalent structure of the two doublets, we infer that an orbital Lambda (\(\Lambda\)) system can be created analogously.

Because even in the presence of strain the leading-order transitions conserve the pseudo-spin,
the spin-conserving transitions to the ES are cyclic if the pseudo-spins inside the KDs are not mixed.
These cycling transitions are used in many platforms for spin readout
\cite{robledo11,delteil14,sukachev17,raha20,appel21}.
Since the coupling to a magnetic field aligned with the crystal axis ($\vec{e}_z$) is diagonal [see Eq.~\eqref{eq:HzEff}]
the pseudo-spins are pure for such a magnetic field.
Therefore, the application of a static magnetic field perfectly aligned with the crystal axis, splitting the spin levels without mixing them,
enables spin readout even in the presence of strain.

\subsection{Pseudo-spin polarization in a highly strained system\label{sec:pol}}
While one possible way to initialize a state is a projective measurement, another approach established in a wide range of platforms is coherent population trapping \cite{gray78,xu08,kelly10,dong12,santori06,togan11,yale13,golter13}.
This approach relies on a Lambda $\Lambda$ system, 
but in the case of the V defect in SiC, all the leading order transitions conserve the pseudo-spin of the KDs.
For this reason, different hyperfine interactions of KDs~\cite{tissot22}, additional fields, or higher-order transition rules are needed to polarize the electron spin.
Higher orders can be investigated using a Schrieffer-Wolff transformation, but are not discussed here for simplicity (see App.~\ref{app:SW} for the case with strain, or  \cite{tissot21a} for the case without strain).

Instead, we briefly outline how the combination of strain and a static magnetic field can be used to set up an optical lambda system.
In particular, we propose to combine \(x\) type strain leading to \(\theta_1 \ne 0\) and a magnetic field in the \(xz\) plane (with non-vanishing \(x,z\) components).
In this case, the KD's Zeeman terms [see Eq.~\eqref{eq:HzEff}] are \(H_{n,\pm,z} =  \mu_B g_{n,\pm}^z \tilde{S}_z B_z + \mu_B g_{n,\pm}^x \tilde{S}_x B_x\)
which is diagonalized by the states
\begin{align}
  \label{eq:Bxεx}
  \ket{n,\pm,\tilde{\sigma}} = & \, \cos(\phi_{n,\pm}/2) \ket{n,\pm,\sigma} \notag \\
  & + \sigma \sin(\phi_{n,\pm}/2) \ket{n,\pm,-\sigma},
\end{align}
with the corresponding eigenvalues, 
\begin{equation}
E_{n,\pm,\tilde{\sigma}}^z = \sigma \mu_B g_{n,\pm}^z B_z \sqrt{1 + (g_{n,\pm}^x B_x)^2/(g_{n,\pm}^z B_z)^2},
\end{equation}
and the angles $\phi_{n,\pm} = \atan\left[ g_{n,\pm}^x B_x / g_{n,\pm}^z B_z \right]$.
With this, an optical Lambda system made up by the two GS \(\ket{1,-,\tilde{\sigma}}\) (\(\sigma = \uparrow,\downarrow\)) and one of the ES becomes feasible.
As an example we calculate the electric dipole matrix element between the GS and the ES \(\ket{2,+,\tilde{\downarrow}}\),
\begin{widetext}
\begin{align}
  \label{eq:lambda}
  \braket{ 1, -, \tilde{\sigma} | H_{\rm el} | 2, +, \tilde{\downarrow} } = \,
& \left[ \cos(\phi_{1,-}/2) \delta_{\sigma,\downarrow} + \sin(\phi_{1,-}/2) \delta_{\sigma,\uparrow} \right] \cos(\phi_{2,+}/2) \braket{ 1, -, \downarrow | H_{\rm el} | 2, +, \downarrow } \notag \\
& + \left[ \sin(\phi_{1,-}/2) \delta_{\sigma,\downarrow} - \cos(\phi_{1,-}/2) \delta_{\sigma,\uparrow} \right] \sin(\phi_{2,+}/2) \braket{ 1, -, \uparrow | H_{\rm el} | 2, +, \uparrow } ,
\end{align}
\end{widetext}
with the spin-conserving matrix elements according to Eq.~\eqref{eq:HelStrain1}.
These selection rules also imply that the corresponding decay processes becomes allowed.
Combined, this enables the preparation of a pseudo-spin state of the GS KD in the presence of strain.

The readout of the qubit discussed in the previous section relied on cyclic transitions.
To make the transitions highly cyclic on demand after using the $\Lambda$ system we can target  $\phi_{n,\pm}=0$ [see Eqs.~\eqref{eq:Bxεx} and \eqref{eq:lambda}]. This can be achieved either by switching the perpendicular component of the magnetic field on and off (e.g., by changing the relative alignment of the magnetic field) or by modulating the perpendicular $g$-tensor component via the strain \cite{ovartchaiyapong14,meesala18,barfuss19} (see Fig.~\ref{fig:gtensor}).
Note that the adiabatic modulation can be sped up by shortcut to adiabadicity approaches like counter adiabatic driving \cite{demirplak03,berry09,guery-odelin19}.

\subsection{Hyperfine interaction in strained KDs\label{sec:hyperfine}}

\begin{figure}[t]
\centering
\includegraphics[width=\linewidth]{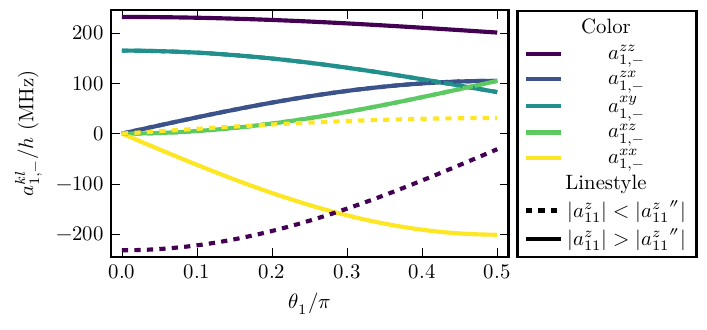}
\caption{\label{fig:hyperfine} Hyperfine tensor elements of the GS KD as functions of the strain mixing angle \(\theta_1\).
The different colors correspond to the different hyperfine tensor elements (see legend).
The solid lines correspond to the case where the orbital hyperfine interaction dominates over the Fermi contact and anisotropic hyperfine interaction and the dashed lines to the reverse.
For most elements both cases align but they significantly differ for $a_{1,-}^{xx}$ and $a_{1,-}^{zz}$.
We use the parameters
$a_{1,-}^x/h = -82.55\,$MHz,
${a_{1,-}^x}'/h = 105\,$MHz,
${a_{1,-}^z}/h = {a_{1,-}^z}'/h = 201\,$MHz ($-31\,$MHz)
${a_{1,-}^z}''/h = -15.5\,$MHz ($100.5\,$MHz)
for the solid (dashed) lines.
}
\end{figure}

After the detailed discussion of the interplay of the electronic structure of TM defects in SiC with strain, we now proceed to the hyperfine structure of the KDs in the presence of strain.
The Hamiltonian of the hyperfine interaction [Eqs.~\eqref{eq:Hhf} and \eqref{eq:Hhf2}] projected onto the strained KDs [Eq.~\eqref{eq:StrainStatesDoublet}] is
\begin{widetext}
\begin{align}
  \label{eq:HhfStrain}
  H_{n,\pm}^{\mathrm{hf}} =
  & \tilde{S}_z \left[a_{n,\pm}^{zz} I_z \pm \frac{a_{n,\pm}^{zx}}{2} \left(e^{i \varphi_n} I_- + e^{-i \varphi_n} I_+\right)\right]
   + \frac{1}{2} \left[e^{- i (1 \pm 1) \varphi_n} \tilde{S}_- \left(a_{n,\pm}^{xy} I_- + a_{n,\pm}^{xx} e^{i \varphi_n} I_+ + a_{n,\pm}^{xz} e^{2 i \varphi_n} I_z\right) + \mathrm{h.c.}\right] ,
\end{align}
\end{widetext}
with the hyperfine coupling constants
\({a_{n,\pm}^{zz} = a_{nn}^z \pm 2 {a_{nn}^z}'' \cos(\theta_n)}\) and \({a_{n,\pm}^{zx} = {a_{nn}^x}' \sin(\theta_n)}\),
\(a_{n,\pm}^{xy} = - a_{nn}^x [1 \mp \cos(\theta_n)]\),
\(a_{n,\pm}^{xz} = {a_{nn}^x}' [1 \pm \cos(\theta_n)]\), and
\(a_{n,\pm}^{xx} = \pm {a_{11}^z}' \sin(\theta_n)\).
We extract the parameters from the literature values (determined at zero strain)~\cite{wolfowicz20,tissot21b,astner22,tissot22} using the following relations by comparing the predicted forms for $\theta_n = 0$.
The average (quarter of the difference) of \(a^{zz}_{n,\pm}\) between the KDs $\pm$ pertaining to the same doublet \(n\) yields
\(a_{nn}^z = \sum_{\sigma=\pm} a^{zz}_{n,\sigma} / 2\)
(\({a_{nn}^z}'' = \sum_{\sigma=\pm} \sigma a^{zz}_{n,\sigma} / 4\)).
The components \(a_{nn}^x\) and \({a_{nn}^x}'\) are fully given by \(a_{n,-}^{xy} / 2\) and \(a_{n,+}^{xz} / 2\), respectively.
We plot the coupling constants of the GS KD \(\ket{1,-,\sigma}\) as a function of the GS strain mixing angle \(\theta_1\) in Fig.~\ref{fig:hyperfine},
where we assume \({a_{nn}^z}' = a_{nn}^z\).
Figure~\ref{fig:hyperfine} shows that due to the symmetry breaking, additional hyperfine elements become non-zero compared with the case of  intact symmetry ($\theta_n = 0$).

While the magnitudes of the different components of the two GS KDs and the lower ES KD are agreed upon within several works \cite{wolfowicz20,tissot21b,tissot22,astner22,hendriks22} and thereby enabled the independent determination of the relative sign of the lowest ES and GS \cite{cilibrizzi23}, the relative sign between the KDs of the same doublet does not have this support.
Therefore, we show in Fig.~\ref{fig:hyperfine} two possible configurations, one for opposite signs (here, $a^{zz}_{1,-} < 0 < a^{zz}_{1,+}$ for $\theta_1 = 0$) where the orbital hyperfine dominates the diagonal interaction and one for the same signs (here, $0 < a^{zz}_{1,-}, a^{zz}_{1,+}$) where the anisotropic hyperfine and Fermi contact interaction are dominant.
The vastly different strain dependencies of the \(zz\) strain coupling elements for the different signs
shows that by measuring this dependence (for example by using a strong, constant magnetic field) it is possible
to determine the relative signs of the hyperfine tensor of the GS KDs without using direct transitions between those.
This would then give us insight into whether the \(|a_{nn}^z|\) or \(|{a^z_{nn}}''|\) is dominant,
where the former stems from the anisotropic and Fermi contact terms and the latter from the orbital angular momentum interacting with the nuclear spin.
Therefore, such a measurement would determine which of these interactions is prevalent.

In the previously proposed nuclear spin-polarization protocol \cite{tissot22}, unstrained samples were considered.
Fundamentally, the different forms of the hyperfine coupling between the KDs were devised to a driven dissipative protocol to polarize the nuclear spin.
In this protocol spin-flipping transitions are driven in combination with spin-conserving decays leading to the polarization of nuclear and electronic spin.
We expect this protocol to be possible if the (\(x,y\)) strain is sufficiently small such that the GS
are dominated by the hyperfine terms \(\propto a_{1,-}^{zz}, a_{1,-}^{xy}\) and the ES by the terms \(\propto a_{2,+}^{zz}, a_{2,+}^{xz}\).
This can be estimated using the strain mixing angles \(\theta_n\) for \(n=1,2\) [see Eq.~\eqref{eq:StrainStatesDoublet}] and Figs.~\ref{fig:hyperfine} and~\ref{fig:splitting}.

On the other hand, for highly (\(x,y\)) strained samples, one has a non-negligible \(\theta_n\).
In this case, one can apply a strong magnetic field along the crystal axis to suppress the hyperfine interaction terms \(\propto \tilde{S}_-,\tilde{S}_+\).
In this case, the quantization axis of the nuclear spin is tilted depending on the KD. This is immediately visible by investigating a single pseudo-spin manifold
\begin{align}  
\label{eq:HhfStrainStrongB}
  & \braket{ n, \pm, \sigma | H | n, \pm, \sigma } \approx E_{n,\pm,\sigma} + (\mu_N g_N B_z + \frac{\sigma}{2} a_{n,\pm}^{zz}) I_z \notag \\ 
  & \qquad \qquad \pm \frac{\sigma}{2} a_{n,\pm}^{zx} [\cos(\varphi_n) I_x - \sin(\varphi_n) I_y], 
\end{align}
with the electronic energy $E_{n,\pm,\sigma}$ and where the magnetic field along \(z\)-direction suppresses the off-diagonal terms.
By first applying a rotation around the \(z\)-axis by the angle \(-\varphi_n\)
and then
a rotation around the \(y\)-axis with the angle \(\pm \atan[(\sigma \mu_N g_N B_z + a_{n,\pm}^{zz})/a_{n,\pm}^{zx}]\),
we diagonalize this manifold, yielding the diagonal hyperfine term in the rotated basis,
\begin{align}\label{eq:HhfStrainStrongBdiag}
  \frac{1}{2} \sqrt{(a_{n,\pm}^{zx})^2 + (2 \mu_N g_N B_z + \sigma a_{n,\pm}^{zz})^2} I_z .
\end{align}
The rotations reflect that the nuclear spin experiences a strain, pseudo-spin, and KD dependent principal axis tilt.
Independent of the pseudo spin, it is possible to drive a (pseudo-spin) conserving transition to an ancillary state (AS) $n,\pm,\sigma$ with a different principal axis tilt incrementally increasing the nuclear-spin polarization.
This enables nuclear polarization.
For simplicity, we will discuss an approach based on purely \(x\)-type strain, i.e. \(\theta_i \ne 0\) and \(\varphi_i = 0\).
Transforming any KDs pseudo-spin $\sigma$ manifold into the diagonal basis of the GS KDs down state (\(1,-,\downarrow\)), we find
that we need to replace $I_z \to \cos(\Delta\varphi) I_z + \sin(\Delta\varphi) I_x$ in Eq.~\eqref{eq:HhfStrainStrongBdiag}
with the relative axis tilt angle \(\Delta \varphi = \atan[(- \mu_N g_N B_z + a_{1,-}^{zz})/a_{1,-}^{zx}] \pm \atan[(\sigma \mu_N g_N B_z + a_{n,\pm}^{zz})/a_{n,\pm}^{zx}]\).
The relative tilt angle $\Delta \varphi$ formalizes that driving to the AS conserving the nuclear-spin state will lead to a nuclear spin precession in the AS until the state decays back to the GS.

This interaction of the nuclear-spin eigenstates of the GS in the AS can be used to polarize the nuclear spin.
A suitable domain for this may be \(\Delta\varphi \ne 0\) but \(\Delta\varphi \ll 1\) where we can approximate the eigenstates of the AS using first-order perturbation theory.
The eigenstates of the ancillary KD pseudo-spin $\sigma$ (in the principal axis system of the GS KD down state) are
\begin{widetext}
\begin{align}
  \label{eq:HhfStrainStrongBdiag-approx}
  \ket{n, \pm, \sigma, m} \approx \ket{n, \pm, \sigma} \left[ \ket{I,m} - \frac{\Delta\varphi}{2} \sqrt{I(I+1) - m(m+1)} \ket{I,m+1} + \frac{\Delta\varphi}{2} \sqrt{I(I+1) - m(m+1)} \ket{I,m-1} \right] .
\end{align}
\end{widetext}
The form of these states shows the possibility to resonantly drive the transitions from the GS \(\sigma,m\) state to the corresponding \(\sigma,m\pm1\) from where the decay mainly occurs to the \(\sigma,m \pm 1\) GS state (if $\Delta \varphi \ll 1$).
Here, the small angle \(\Delta \varphi\) ensures that the main decay does not decrease the nuclear magnetic moment again, while still enabling a resonant drive of the polarizing transition.
This process polarizes the nuclear spin stepwise.
The outlined approach works with the transition to an ES or the second GS,
given that the nuclear transition lines can be resolved.
This is in contrast to the proposed zero-strain protocol, where a pseudo-spin flipping transition in combination with the correct polarization renders this requirement unnecessary.

While we focused for concreteness on one additional example including strain in this work,
we note that combining our model  with the general approach outlined in \cite{tissot22},
protocols optimized for different scenarios can be developed that are best matched to the technical setup.

\section{Conclusion\label{sec:conclusion}}
We studied the influence of strain on transition-metal defects in SiC, focusing on a particularly promising center for quantum technology applications, the substitutional vanadium defect at the $k$ site of 4H-SiC.
We found that using strain enables the engineering of the electronic \(g\)-tensor and optical selection rules, thereby opening the possibility for strain-controlled manipulation (microwave gates) within the KDs,
as well as \(\Lambda\) and V optical three-level setups where both branches can be driven using the same polarization.
By combining strain and magnetic fields, we showed a path towards engineering \(\Lambda\) systems for the pseudo-spin states of the KD,
thus enabling further prospects such as state preparation within the GS KD.
We also discussed the prospect of state readout of strained defects using cycling transitions.

Furthermore, we showed the influence of strain on the hyperfine interaction within the KDs.
Here, we found that the previously proposed polarization protocol is likely not applicable anymore for strongly strained samples.
Therefore, we discussed one example of an application of our theory where it is straightforward to find different polarization schemes even in the presence of strain.
%
A natural next step would be to exploit our theoretical insights in future experiments.

\begin{acknowledgments}
  We acknowledge funding from the European Union’s Horizon 2020 research and innovation programme under Grant Agreement No.~862721 (QuanTELCO).
  B.T. and G.B. additionally acknowledge funding from the German Federal Ministry of Education and Research (BMBF) under the Grant Agreement No.~13N16212 (SPINNING).
  A.G. acknowledges the support by the National Excellence Program for the project of Quantum-coherent materials (NKFIH Grant No.~KKP129866) as well as by the Ministry of Culture and Innovation and the National Research, Development and Innovation Office within the Quantum Information National Laboratory of Hungary (Grant No.~2022-2.1.1-NL-2022-00004). A.G. additionally acknowledges the high-performance computational resources provided by KIFU (Governmental Agency for IT Development) institute of Hungary and the European Commission for the project QuMicro (Grant No.~101046911).
\end{acknowledgments}

\appendix
\section{DFT calculation methods\label{app:DFT}}

We model the vanadium defect embedded in a 128-atom 4H-SiC supercell. Its electronic structure is calculated using the plane-wave based Vienna Ab-initio Simulation Package (VASP)~\cite{VASP1, VASP2, VASP3, VASP4}, with the $\Gamma$-point approximation for the k-point sampling. The plane wave cutoff is set to 420\,eV and PAW method~\cite{PAW} is used for the core electrons. We apply DFT using the hybrid exchange functional of Heyd, Scuseria, and Ernzerhof (HSE06)~\cite{HSE06} with on-site  correction (DFT+U) according to the Dudarev-approach~\cite{Dudarev_1998}, where the d-orbitals of the vanadium atom is effected by $U=-2.5\,$eV~\cite{Ivady_2013}. The atomic configurations are relaxed to forces smaller than 0.01\,eV/\AA. Excited state electronic configurations are calculated with the constrained occupation $\Delta$-SCF method~\cite{Gali_2009}. The methods for applying strain is discussed in Ref.~\cite{udvarhelyi18}. We note that the local defect structure is relaxed within the strain constraint applied to the lattice. 

The results of this simulation are discussed in the main text and we additionally provided all slopes extracted for the coupling to strain in Tab.~\ref{tab:DFT_strain}.
In this table ${s_{nm}^{y}}^{(\prime)}$ takes the same role as ${s_{nm}^{x}}^{(\prime)}$ [see Eq.~\eqref{eq:Hstrain}]
but takes into account that we cannot assume them to be the same \emph{a priori} within the DFT calculation.
We encode the slope of the ES and GS level splitting in $s_{22}^z, {s_{22}^z}'$ by choosing $s_{11}^z, {s_{11}^z}'=0$ as we discuss in the main text.
\begin{table}[]
    \centering
    \begin{tabular}{ccc}
     symmetry & parameter    & calculated value (eV/strain) \\\hline 
      \multirow{4}{*}{$E_x$}    &$s_{11}^{x}$   & 1.04(1)\\
      &$s_{22}^{x}$   & 0.56(5)\\
      &${s_{11}^{x}}'$   & 0.94(2)\\
      &${s_{22}^{x}}'$   & 0.85(2)\\\hline
      \multirow{4}{*}{$E_y$}&$s_{11}^{y}$   & 1.037(2)\\
      &$s_{22}^{y}$   & 0.582(2)\\
      &${s_{11}^{y}}'$   & 0.958(4)\\
      &${s_{22}^{y}}'$   & 0.84(1)\\\hline
      \multirow{2}{*}{$A_1$}&$s_{22}^{z}$   & 1.9(1)\\
      &${s_{22}^{z}}'$   & 1.26(8)
    \end{tabular}
    \caption{Calculated strain-orbital coupling coefficients extracted from linear perturbation model of the orbital level splitting and ZPL energies from DFT. Couplings of $E_x$ and $E_y$ perturbations show the same coupling strengths within computational accuracy. Standard deviations are extracted from the linear fit and are given in parenthesis for the last meaningful digit.
    We note that the $s_{22}^{z},{s_{22}^{z}}'$ values correspond to the slope of the ES-GS splitting where we chose $s_{11}^{z},{s_{11}^{z}}'=0$.
    }
    \label{tab:DFT_strain}
\end{table}

\section{Crystal field eigenstates and signs of the strain coupling constants\label{app:strainsign}}
\label{sec:orgcba8a69}
Inside the \(d\) orbital projections the crystal eigenstates are given as
\begin{equation}
  \label{eq:HcrStates}
  \begin{aligned}
  \ket{\pm_1} = & \cos(\phi) \ket{\pm 1} \mp \sin(\phi) \ket{\mp 2} , \\
  \ket{\pm_2} = & - \sin(\phi) \ket{\pm 1} \mp \cos(\phi) \ket{\mp 2},
  \end{aligned}
\end{equation}
where the crystal mixing angle \(\phi\) describes the admixture of states that transform equally under \(C_{3v}\).

In the absence of spin-orbit coupling the doublet states split due to \(x,y\)-strain, we use this to determine the sign of the strain coupling constants.
The eigenvectors for purely \(\epsilon^x_{nm}\) strain are (within the doublet projection) given by
\begin{align}
  \frac{\ket{+_1} \pm \ket{-_1}}{\sqrt{2}} =\, & \cos(\phi) ( \ket{+ 1} \pm \ket{-1} ) \notag \\
  \label{eq:purestrainstates}
  & - \sin(\phi) ( \ket{- 2} \mp \ket{+ 2} ), \\
  \frac{\ket{+_2} \pm \ket{-_2}}{\sqrt{2}} =\, & - \sin(\phi) ( \ket{+ 1} \pm \ket{-1} ) \notag \\
  \label{eq:purestrainstates2}
  & - \cos(\phi) ( \ket{- 2} \mp \ket{+ 2} ),
\end{align}
with the eigenvalues \(\epsilon_n \pm \epsilon_{nn}^x\).
We note the parallel of these pairs of states to the cubic harmonics
\(\ket{x^2 - y^2} = \frac{1}{\sqrt{2}} \left( \ket{+2} + \ket{-2} \right)\),
\(\ket{xy} = \frac{i}{\sqrt{2}} \left( - \ket{+2} + \ket{-2} \right)\),
\(\ket{xz} = \frac{1}{\sqrt{2}} \left( - \ket{+1} + \ket{-1} \right)\),
\(\ket{yz} = \frac{i}{\sqrt{2}} \left( \ket{+1} + \ket{-1} \right)\),
and
\(\ket{z^2} = \ket{0}\) .
With this we find that the strain eigenstates are proportional to the distinct sets of cubic harmonics
\(\ket{+_n} + \ket{-_n} \propto \ket{yz}, \ket{xy}\) and
\(\ket{+_n} - \ket{-_n} \propto \ket{xz}, \ket{x^2 - y^2}\).

With this we can obtain the sign of the coupling from the DFT simulation without spin by comparing the projection on the cubic harmonics (of the \(d\)-orbital).
We find that for \(\epsilon_{xz} = 0.01\) in the GS (ES) the lower energy state is  mainly \(\ket{xz}\) (\(\ket{xy}\)), i.e. the \(\ket{+_1} - \ket{-_1}\) (\(\ket{+_1} + \ket{-_1}\)) state,
such that \({s_{11}^x} > 0\) (\({s_{22}^x} < 0\)).
Analogously, for \(\epsilon_{yy} - \epsilon_{xx} = 0.01\) in the GS (ES) the lower energy state is  mainly \(\ket{xz}\) (\(\ket{xy}\)), i.e. the \(\ket{+_1} - \ket{-_1}\) (\(\ket{+_2} + \ket{+_2}\)) state,
such that \({s_{11}^x}' > 0\) (\({s_{22}^x}' < 0\)).

Due to the known transformation properties of the states from the literature (using the difference in the hyperfine tensor)
we assign the lower KD of the GS in the absence of strain to \(\Gamma_4\) and in the ES to \(\Gamma_{5/6}\) \cite{wolfowicz20,tissot21b,tissot22,astner22,cilibrizzi23}.
To accommodate this in the model, we use that for the vanadium  defect in the \(k\) site in 4H-SiC,
\(\lambda_{11}^z > 0\) and \(\lambda_{22}^z < 0\).

\section{Higher-order effects\label{app:SW}}
To understand higher order effects we treat a purely \(x\)-type strain using a Schrieffer-Wolff transformation \cite{bravyi11}.
To this end, we perturbatively take block off-diagonal elements of the spin-orbit and strain Hamiltonians (together) into account.
We do the following calculations in the basis where the leading-order doublet Hamiltonians \(P_n H_{\mathrm{ezf}} P_n\) given in Eq.~\eqref{eq:StrainStatesDoublet} are diagonalized, such that we can afterwards directly study the corrections affecting the KDs.
Then we use the transformation $U=\exp(-S)$
and, within first-order perturbation theory,
\begin{align}
  \label{eq:SW1}
  S = \sum_{n \ne m} P_n (H_{\mathrm{so}} + H_{\mathrm{st}}) P_m / (\epsilon_n - \epsilon_m),
\end{align}
where we directly neglect spin-orbit and strain terms in the denominator as they are part of the higher (neglected) orders.
The corrections to the zero-field energies are then given by \(\frac{1}{2} \comm{S}{H_{\mathrm{ezf}}}\) which is block-diagonal in the KDs and corrects their energies by
\begin{widetext}
\begin{align}
  \label{eq:EnergieCorrection}
  {E_{n,-}}' = & (-1)^n \frac{{\lambda_{12}^z}^2/4 + {\lambda_{12}^x}^2 \sin^2(\theta_n/2) + {\epsilon_{12}^x}^2}{\epsilon_2 - \epsilon_1} - \frac{{\lambda_{i3}^x}^2 \cos^2(\theta_n/2) + {\epsilon_{i3}^x}^2 (1 + \sin(\theta_n))}{\epsilon_3 - \epsilon_1} , \\
  {E_{n,+}}' = & (-1)^n \frac{{\lambda_{12}^z}^2/4 + {\lambda_{12}^x}^2 \cos^2(\theta_n/2) + {\epsilon_{12}^x}^2}{\epsilon_2 - \epsilon_1} - \frac{{\lambda_{i3}^x}^2 \sin^2(\theta_n/2) + {\epsilon_{i3}^x}^2 (1 - \sin(\theta_n))}{\epsilon_3 - \epsilon_1} , \\
  {E_3}' = & \frac{{\lambda_{23}^x}^2 + 2 {\epsilon_{23}^x}^2}{\epsilon_3 - \epsilon_2} + \frac{{\lambda_{13}^x}^2 + 2 {\epsilon_{13}^x}^2}{\epsilon_3 - \epsilon_1} ,
\end{align}
where \(n = 1,2\).

In addition to this, the corresponding corrections of the remaining parts \(h\) of the full Hamiltonian can be calculated as \(\comm{S}{h}\).
For instance this corrects the coupling to a magnetic field along the crystal axis \(\vec{e}_z\) projected onto the KDs as
\begin{align}
  \label{eq:Hzcorr}
  {H_{n,-}^z}' = & \left[ (-1)^n \frac{2 \lambda_{12}^z}{\epsilon_2 - \epsilon_1} S_z - \frac{4 \lambda_{12}^x \sin^2(\theta_n/2)}{\epsilon_2 - \epsilon_1} S_x \right] r_{12}^z \mu_B B_z , \\
  {H_{n,+}^z}' = & \left[ (-1)^n \frac{2 \lambda_{12}^z}{\epsilon_2 - \epsilon_1} S_z - \frac{4 \lambda_{12}^x \cos^2(\theta_n/2)}{\epsilon_2 - \epsilon_1} S_x \right] r_{12}^z \mu_B B_z , \\
  {H_{3}^z}' = & 0,
\end{align}
\end{widetext}
where one can neglect the off-diagonal matrix elements considering that they are suppressed by the leading-order term \(g_s S_z\), as we expect for \(g_s \gg r_{12}^z\).
While in this article we focus on providing the straight-forward recipe to calculate higher-order terms for simplicity, previous work takes higher-order effects in the spin-orbit coupling only (without strain) into account~\cite{tissot21a,tissot21b,tissot22}.
Analogously, expressions for other magnetic-field directions and parts of the Hamiltonian can be calculated using \(S\) but are omitted here.

\bibliography{refs.bib}

\begin{thebibliography}{78}%
\makeatletter
\providecommand \@ifxundefined [1]{%
 \@ifx{#1\undefined}
}%
\providecommand \@ifnum [1]{%
 \ifnum #1\expandafter \@firstoftwo
 \else \expandafter \@secondoftwo
 \fi
}%
\providecommand \@ifx [1]{%
 \ifx #1\expandafter \@firstoftwo
 \else \expandafter \@secondoftwo
 \fi
}%
\providecommand \natexlab [1]{#1}%
\providecommand \enquote  [1]{``#1''}%
\providecommand \bibnamefont  [1]{#1}%
\providecommand \bibfnamefont [1]{#1}%
\providecommand \citenamefont [1]{#1}%
\providecommand \href@noop [0]{\@secondoftwo}%
\providecommand \href [0]{\begingroup \@sanitize@url \@href}%
\providecommand \@href[1]{\@@startlink{#1}\@@href}%
\providecommand \@@href[1]{\endgroup#1\@@endlink}%
\providecommand \@sanitize@url [0]{\catcode `\\12\catcode `\$12\catcode
  `\&12\catcode `\#12\catcode `\^12\catcode `\_12\catcode `\%12\relax}%
\providecommand \@@startlink[1]{}%
\providecommand \@@endlink[0]{}%
\providecommand \url  [0]{\begingroup\@sanitize@url \@url }%
\providecommand \@url [1]{\endgroup\@href {#1}{\urlprefix }}%
\providecommand \urlprefix  [0]{URL }%
\providecommand \Eprint [0]{\href }%
\providecommand \doibase [0]{https://doi.org/}%
\providecommand \selectlanguage [0]{\@gobble}%
\providecommand \bibinfo  [0]{\@secondoftwo}%
\providecommand \bibfield  [0]{\@secondoftwo}%
\providecommand \translation [1]{[#1]}%
\providecommand \BibitemOpen [0]{}%
\providecommand \bibitemStop [0]{}%
\providecommand \bibitemNoStop [0]{.\EOS\space}%
\providecommand \EOS [0]{\spacefactor3000\relax}%
\providecommand \BibitemShut  [1]{\csname bibitem#1\endcsname}%
\let\auto@bib@innerbib\@empty
\bibitem [{\citenamefont {Aharonovich}\ \emph {et~al.}(2016)\citenamefont
  {Aharonovich}, \citenamefont {Englund},\ and\ \citenamefont
  {Toth}}]{aharonovich16}%
  \BibitemOpen
  \bibfield  {author} {\bibinfo {author} {\bibfnamefont {I.}~\bibnamefont
  {Aharonovich}}, \bibinfo {author} {\bibfnamefont {D.}~\bibnamefont
  {Englund}},\ and\ \bibinfo {author} {\bibfnamefont {M.}~\bibnamefont
  {Toth}},\ }\bibfield  {title} {\bibinfo {title} {Solid-state single-photon
  emitters},\ }\href {https://doi.org/10.1038/nphoton.2016.186} {\bibfield
  {journal} {\bibinfo  {journal} {Nat. Photonics}\ }\textbf {\bibinfo {volume}
  {10}},\ \bibinfo {pages} {631} (\bibinfo {year} {2016})}\BibitemShut
  {NoStop}%
\bibitem [{\citenamefont {Heshami}\ \emph {et~al.}(2016)\citenamefont
  {Heshami}, \citenamefont {England}, \citenamefont {Humphreys}, \citenamefont
  {Bustard}, \citenamefont {Acosta}, \citenamefont {Nunn},\ and\ \citenamefont
  {Sussman}}]{heshami16}%
  \BibitemOpen
  \bibfield  {author} {\bibinfo {author} {\bibfnamefont {K.}~\bibnamefont
  {Heshami}}, \bibinfo {author} {\bibfnamefont {D.~G.}\ \bibnamefont
  {England}}, \bibinfo {author} {\bibfnamefont {P.~C.}\ \bibnamefont
  {Humphreys}}, \bibinfo {author} {\bibfnamefont {P.~J.}\ \bibnamefont
  {Bustard}}, \bibinfo {author} {\bibfnamefont {V.~M.}\ \bibnamefont {Acosta}},
  \bibinfo {author} {\bibfnamefont {J.}~\bibnamefont {Nunn}},\ and\ \bibinfo
  {author} {\bibfnamefont {B.~J.}\ \bibnamefont {Sussman}},\ }\bibfield
  {title} {\bibinfo {title} {Quantum memories: emerging applications and recent
  advances},\ }\href {https://doi.org/10.1080/09500340.2016.1148212} {\bibfield
   {journal} {\bibinfo  {journal} {J. Mod. Opt.}\ }\textbf {\bibinfo {volume}
  {63}},\ \bibinfo {pages} {2005} (\bibinfo {year} {2016})}\BibitemShut
  {NoStop}%
\bibitem [{\citenamefont {Awschalom}\ \emph {et~al.}(2021)\citenamefont
  {Awschalom}, \citenamefont {Berggren}, \citenamefont {Bernien}, \citenamefont
  {Bhave}, \citenamefont {Carr}, \citenamefont {Davids}, \citenamefont
  {Economou}, \citenamefont {Englund}, \citenamefont {Faraon}, \citenamefont
  {Fejer}, \citenamefont {Guha}, \citenamefont {Gustafsson}, \citenamefont
  {Hu}, \citenamefont {Jiang}, \citenamefont {Kim}, \citenamefont {Korzh},
  \citenamefont {Kumar}, \citenamefont {Kwiat}, \citenamefont {Lončar},
  \citenamefont {Lukin}, \citenamefont {Miller}, \citenamefont {Monroe},
  \citenamefont {Nam}, \citenamefont {Narang}, \citenamefont {Orcutt},
  \citenamefont {Raymer}, \citenamefont {Safavi-Naeini}, \citenamefont
  {Spiropulu}, \citenamefont {Srinivasan}, \citenamefont {Sun}, \citenamefont
  {Vučković}, \citenamefont {Waks}, \citenamefont {Walsworth}, \citenamefont
  {Weiner},\ and\ \citenamefont {Zhang}}]{awschalom21}%
  \BibitemOpen
  \bibfield  {author} {\bibinfo {author} {\bibfnamefont {D.}~\bibnamefont
  {Awschalom}}, \bibinfo {author} {\bibfnamefont {K.~K.}\ \bibnamefont
  {Berggren}}, \bibinfo {author} {\bibfnamefont {H.}~\bibnamefont {Bernien}},
  \bibinfo {author} {\bibfnamefont {S.}~\bibnamefont {Bhave}}, \bibinfo
  {author} {\bibfnamefont {L.~D.}\ \bibnamefont {Carr}}, \bibinfo {author}
  {\bibfnamefont {P.}~\bibnamefont {Davids}}, \bibinfo {author} {\bibfnamefont
  {S.~E.}\ \bibnamefont {Economou}}, \bibinfo {author} {\bibfnamefont
  {D.}~\bibnamefont {Englund}}, \bibinfo {author} {\bibfnamefont
  {A.}~\bibnamefont {Faraon}}, \bibinfo {author} {\bibfnamefont
  {M.}~\bibnamefont {Fejer}}, \bibinfo {author} {\bibfnamefont
  {S.}~\bibnamefont {Guha}}, \bibinfo {author} {\bibfnamefont {M.~V.}\
  \bibnamefont {Gustafsson}}, \bibinfo {author} {\bibfnamefont
  {E.}~\bibnamefont {Hu}}, \bibinfo {author} {\bibfnamefont {L.}~\bibnamefont
  {Jiang}}, \bibinfo {author} {\bibfnamefont {J.}~\bibnamefont {Kim}}, \bibinfo
  {author} {\bibfnamefont {B.}~\bibnamefont {Korzh}}, \bibinfo {author}
  {\bibfnamefont {P.}~\bibnamefont {Kumar}}, \bibinfo {author} {\bibfnamefont
  {P.~G.}\ \bibnamefont {Kwiat}}, \bibinfo {author} {\bibfnamefont
  {M.}~\bibnamefont {Lončar}}, \bibinfo {author} {\bibfnamefont {M.~D.}\
  \bibnamefont {Lukin}}, \bibinfo {author} {\bibfnamefont {D.~A.}\ \bibnamefont
  {Miller}}, \bibinfo {author} {\bibfnamefont {C.}~\bibnamefont {Monroe}},
  \bibinfo {author} {\bibfnamefont {S.~W.}\ \bibnamefont {Nam}}, \bibinfo
  {author} {\bibfnamefont {P.}~\bibnamefont {Narang}}, \bibinfo {author}
  {\bibfnamefont {J.~S.}\ \bibnamefont {Orcutt}}, \bibinfo {author}
  {\bibfnamefont {M.~G.}\ \bibnamefont {Raymer}}, \bibinfo {author}
  {\bibfnamefont {A.~H.}\ \bibnamefont {Safavi-Naeini}}, \bibinfo {author}
  {\bibfnamefont {M.}~\bibnamefont {Spiropulu}}, \bibinfo {author}
  {\bibfnamefont {K.}~\bibnamefont {Srinivasan}}, \bibinfo {author}
  {\bibfnamefont {S.}~\bibnamefont {Sun}}, \bibinfo {author} {\bibfnamefont
  {J.}~\bibnamefont {Vučković}}, \bibinfo {author} {\bibfnamefont
  {E.}~\bibnamefont {Waks}}, \bibinfo {author} {\bibfnamefont {R.}~\bibnamefont
  {Walsworth}}, \bibinfo {author} {\bibfnamefont {A.~M.}\ \bibnamefont
  {Weiner}},\ and\ \bibinfo {author} {\bibfnamefont {Z.}~\bibnamefont
  {Zhang}},\ }\bibfield  {title} {\bibinfo {title} {Development of quantum
  interconnects (quics) for next-generation information technologies},\ }\href
  {https://doi.org/10.1103/prxquantum.2.017002} {\bibfield  {journal} {\bibinfo
   {journal} {PRX Quantum}\ }\textbf {\bibinfo {volume} {2}},\ \bibinfo {pages}
  {017002} (\bibinfo {year} {2021})}\BibitemShut {NoStop}%
\bibitem [{\citenamefont {He}\ \emph {et~al.}(1993)\citenamefont {He},
  \citenamefont {Manson},\ and\ \citenamefont {Fisk}}]{he93}%
  \BibitemOpen
  \bibfield  {author} {\bibinfo {author} {\bibfnamefont {X.-F.}\ \bibnamefont
  {He}}, \bibinfo {author} {\bibfnamefont {N.~B.}\ \bibnamefont {Manson}},\
  and\ \bibinfo {author} {\bibfnamefont {P.~T.~H.}\ \bibnamefont {Fisk}},\
  }\bibfield  {title} {\bibinfo {title} {Paramagnetic resonance of photoexcited
  {N}-{V} defects in diamond. {I}. level anticrossing in the $^3${A} ground
  state},\ }\href {https://doi.org/10.1103/physrevb.47.8809} {\bibfield
  {journal} {\bibinfo  {journal} {Phys. Rev. B}\ }\textbf {\bibinfo {volume}
  {47}},\ \bibinfo {pages} {8809} (\bibinfo {year} {1993})}\BibitemShut
  {NoStop}%
\bibitem [{\citenamefont {Gaebel}\ \emph {et~al.}(2006)\citenamefont {Gaebel},
  \citenamefont {Domhan}, \citenamefont {Popa}, \citenamefont {Wittmann},
  \citenamefont {Neumann}, \citenamefont {Jelezko}, \citenamefont {Rabeau},
  \citenamefont {Stavrias}, \citenamefont {Greentree}, \citenamefont {Prawer},
  \citenamefont {Meijer}, \citenamefont {Twamley}, \citenamefont {Hemmer},\
  and\ \citenamefont {Wrachtrup}}]{gaebel06}%
  \BibitemOpen
  \bibfield  {author} {\bibinfo {author} {\bibfnamefont {T.}~\bibnamefont
  {Gaebel}}, \bibinfo {author} {\bibfnamefont {M.}~\bibnamefont {Domhan}},
  \bibinfo {author} {\bibfnamefont {I.}~\bibnamefont {Popa}}, \bibinfo {author}
  {\bibfnamefont {C.}~\bibnamefont {Wittmann}}, \bibinfo {author}
  {\bibfnamefont {P.}~\bibnamefont {Neumann}}, \bibinfo {author} {\bibfnamefont
  {F.}~\bibnamefont {Jelezko}}, \bibinfo {author} {\bibfnamefont {J.~R.}\
  \bibnamefont {Rabeau}}, \bibinfo {author} {\bibfnamefont {N.}~\bibnamefont
  {Stavrias}}, \bibinfo {author} {\bibfnamefont {A.~D.}\ \bibnamefont
  {Greentree}}, \bibinfo {author} {\bibfnamefont {S.}~\bibnamefont {Prawer}},
  \bibinfo {author} {\bibfnamefont {J.}~\bibnamefont {Meijer}}, \bibinfo
  {author} {\bibfnamefont {J.}~\bibnamefont {Twamley}}, \bibinfo {author}
  {\bibfnamefont {P.~R.}\ \bibnamefont {Hemmer}},\ and\ \bibinfo {author}
  {\bibfnamefont {J.}~\bibnamefont {Wrachtrup}},\ }\bibfield  {title} {\bibinfo
  {title} {Room-temperature coherent coupling of single spins in diamond},\
  }\href {https://doi.org/10.1038/nphys318} {\bibfield  {journal} {\bibinfo
  {journal} {Nature Phys.}\ }\textbf {\bibinfo {volume} {2}},\ \bibinfo {pages}
  {408} (\bibinfo {year} {2006})}\BibitemShut {NoStop}%
\bibitem [{\citenamefont {Childress}\ \emph {et~al.}(2006)\citenamefont
  {Childress}, \citenamefont {Dutt}, \citenamefont {Taylor}, \citenamefont
  {Zibrov}, \citenamefont {Jelezko}, \citenamefont {Wrachtrup}, \citenamefont
  {Hemmer},\ and\ \citenamefont {Lukin}}]{childress06}%
  \BibitemOpen
  \bibfield  {author} {\bibinfo {author} {\bibfnamefont {L.}~\bibnamefont
  {Childress}}, \bibinfo {author} {\bibfnamefont {M.~V.~G.}\ \bibnamefont
  {Dutt}}, \bibinfo {author} {\bibfnamefont {J.~M.}\ \bibnamefont {Taylor}},
  \bibinfo {author} {\bibfnamefont {A.~S.}\ \bibnamefont {Zibrov}}, \bibinfo
  {author} {\bibfnamefont {F.}~\bibnamefont {Jelezko}}, \bibinfo {author}
  {\bibfnamefont {J.}~\bibnamefont {Wrachtrup}}, \bibinfo {author}
  {\bibfnamefont {P.~R.}\ \bibnamefont {Hemmer}},\ and\ \bibinfo {author}
  {\bibfnamefont {M.~D.}\ \bibnamefont {Lukin}},\ }\bibfield  {title} {\bibinfo
  {title} {Coherent dynamics of coupled electron and nuclear spin qubits in
  diamond},\ }\href {https://doi.org/10.1126/science.1131871} {\bibfield
  {journal} {\bibinfo  {journal} {Science}\ }\textbf {\bibinfo {volume}
  {314}},\ \bibinfo {pages} {281} (\bibinfo {year} {2006})}\BibitemShut
  {NoStop}%
\bibitem [{\citenamefont {Santori}\ \emph {et~al.}(2006)\citenamefont
  {Santori}, \citenamefont {Tamarat}, \citenamefont {Neumann}, \citenamefont
  {Wrachtrup}, \citenamefont {Fattal}, \citenamefont {Beausoleil},
  \citenamefont {Rabeau}, \citenamefont {Olivero}, \citenamefont {Greentree},
  \citenamefont {Prawer}, \citenamefont {Jelezko},\ and\ \citenamefont
  {Hemmer}}]{santori06}%
  \BibitemOpen
  \bibfield  {author} {\bibinfo {author} {\bibfnamefont {C.}~\bibnamefont
  {Santori}}, \bibinfo {author} {\bibfnamefont {P.}~\bibnamefont {Tamarat}},
  \bibinfo {author} {\bibfnamefont {P.}~\bibnamefont {Neumann}}, \bibinfo
  {author} {\bibfnamefont {J.}~\bibnamefont {Wrachtrup}}, \bibinfo {author}
  {\bibfnamefont {D.}~\bibnamefont {Fattal}}, \bibinfo {author} {\bibfnamefont
  {R.~G.}\ \bibnamefont {Beausoleil}}, \bibinfo {author} {\bibfnamefont
  {J.}~\bibnamefont {Rabeau}}, \bibinfo {author} {\bibfnamefont
  {P.}~\bibnamefont {Olivero}}, \bibinfo {author} {\bibfnamefont {A.~D.}\
  \bibnamefont {Greentree}}, \bibinfo {author} {\bibfnamefont {S.}~\bibnamefont
  {Prawer}}, \bibinfo {author} {\bibfnamefont {F.}~\bibnamefont {Jelezko}},\
  and\ \bibinfo {author} {\bibfnamefont {P.}~\bibnamefont {Hemmer}},\
  }\bibfield  {title} {\bibinfo {title} {Coherent population trapping of single
  spins in diamond under optical excitation},\ }\href
  {https://doi.org/10.1103/physrevlett.97.247401} {\bibfield  {journal}
  {\bibinfo  {journal} {Phys. Rev. Lett.}\ }\textbf {\bibinfo {volume} {97}},\
  \bibinfo {pages} {247401} (\bibinfo {year} {2006})}\BibitemShut {NoStop}%
\bibitem [{\citenamefont {Gali}\ \emph {et~al.}(2008)\citenamefont {Gali},
  \citenamefont {Fyta},\ and\ \citenamefont {Kaxiras}}]{gali08}%
  \BibitemOpen
  \bibfield  {author} {\bibinfo {author} {\bibfnamefont {A.}~\bibnamefont
  {Gali}}, \bibinfo {author} {\bibfnamefont {M.}~\bibnamefont {Fyta}},\ and\
  \bibinfo {author} {\bibfnamefont {E.}~\bibnamefont {Kaxiras}},\ }\bibfield
  {title} {\bibinfo {title} {Ab initio supercell calculations on
  nitrogen-vacancy center in diamond: Electronic structure and hyperfine
  tensors},\ }\href {https://doi.org/10.1103/physrevb.77.155206} {\bibfield
  {journal} {\bibinfo  {journal} {Phys. Rev. B}\ }\textbf {\bibinfo {volume}
  {77}},\ \bibinfo {pages} {155206} (\bibinfo {year} {2008})}\BibitemShut
  {NoStop}%
\bibitem [{\citenamefont {Felton}\ \emph {et~al.}(2009)\citenamefont {Felton},
  \citenamefont {Edmonds}, \citenamefont {Newton}, \citenamefont {Martineau},
  \citenamefont {Fisher}, \citenamefont {Twitchen},\ and\ \citenamefont
  {Baker}}]{felton09}%
  \BibitemOpen
  \bibfield  {author} {\bibinfo {author} {\bibfnamefont {S.}~\bibnamefont
  {Felton}}, \bibinfo {author} {\bibfnamefont {A.~M.}\ \bibnamefont {Edmonds}},
  \bibinfo {author} {\bibfnamefont {M.~E.}\ \bibnamefont {Newton}}, \bibinfo
  {author} {\bibfnamefont {P.~M.}\ \bibnamefont {Martineau}}, \bibinfo {author}
  {\bibfnamefont {D.}~\bibnamefont {Fisher}}, \bibinfo {author} {\bibfnamefont
  {D.~J.}\ \bibnamefont {Twitchen}},\ and\ \bibinfo {author} {\bibfnamefont
  {J.~M.}\ \bibnamefont {Baker}},\ }\bibfield  {title} {\bibinfo {title}
  {Hyperfine interaction in the ground state of the negatively charged nitrogen
  vacancy center in diamond},\ }\href
  {https://doi.org/10.1103/physrevb.79.075203} {\bibfield  {journal} {\bibinfo
  {journal} {Phys. Rev. B}\ }\textbf {\bibinfo {volume} {79}},\ \bibinfo
  {pages} {075203} (\bibinfo {year} {2009})}\BibitemShut {NoStop}%
\bibitem [{\citenamefont {Maze}\ \emph {et~al.}(2011)\citenamefont {Maze},
  \citenamefont {Gali}, \citenamefont {Togan}, \citenamefont {Chu},
  \citenamefont {Trifonov}, \citenamefont {Kaxiras},\ and\ \citenamefont
  {Lukin}}]{maze11}%
  \BibitemOpen
  \bibfield  {author} {\bibinfo {author} {\bibfnamefont {J.~R.}\ \bibnamefont
  {Maze}}, \bibinfo {author} {\bibfnamefont {A.}~\bibnamefont {Gali}}, \bibinfo
  {author} {\bibfnamefont {E.}~\bibnamefont {Togan}}, \bibinfo {author}
  {\bibfnamefont {Y.}~\bibnamefont {Chu}}, \bibinfo {author} {\bibfnamefont
  {A.}~\bibnamefont {Trifonov}}, \bibinfo {author} {\bibfnamefont
  {E.}~\bibnamefont {Kaxiras}},\ and\ \bibinfo {author} {\bibfnamefont {M.~D.}\
  \bibnamefont {Lukin}},\ }\bibfield  {title} {\bibinfo {title} {Properties of
  nitrogen-vacancy centers in diamond: the group theoretic approach},\ }\href
  {https://doi.org/10.1088/1367-2630/13/2/025025} {\bibfield  {journal}
  {\bibinfo  {journal} {New J. Phys.}\ }\textbf {\bibinfo {volume} {13}},\
  \bibinfo {pages} {025025} (\bibinfo {year} {2011})}\BibitemShut {NoStop}%
\bibitem [{\citenamefont {Fuchs}\ \emph {et~al.}(2011)\citenamefont {Fuchs},
  \citenamefont {Burkard}, \citenamefont {Klimov},\ and\ \citenamefont
  {Awschalom}}]{fuchs11}%
  \BibitemOpen
  \bibfield  {author} {\bibinfo {author} {\bibfnamefont {G.~D.}\ \bibnamefont
  {Fuchs}}, \bibinfo {author} {\bibfnamefont {G.}~\bibnamefont {Burkard}},
  \bibinfo {author} {\bibfnamefont {P.~V.}\ \bibnamefont {Klimov}},\ and\
  \bibinfo {author} {\bibfnamefont {D.~D.}\ \bibnamefont {Awschalom}},\
  }\bibfield  {title} {\bibinfo {title} {{A quantum memory intrinsic to single
  nitrogen{\textendash}vacancy centres in diamond}},\ }\href
  {https://doi.org/10.1038/nphys2026} {\bibfield  {journal} {\bibinfo
  {journal} {Nature Phys.}\ }\textbf {\bibinfo {volume} {7}},\ \bibinfo {pages}
  {789} (\bibinfo {year} {2011})}\BibitemShut {NoStop}%
\bibitem [{\citenamefont {Togan}\ \emph {et~al.}(2011)\citenamefont {Togan},
  \citenamefont {Chu}, \citenamefont {Imamoglu},\ and\ \citenamefont
  {Lukin}}]{togan11}%
  \BibitemOpen
  \bibfield  {author} {\bibinfo {author} {\bibfnamefont {E.}~\bibnamefont
  {Togan}}, \bibinfo {author} {\bibfnamefont {Y.}~\bibnamefont {Chu}}, \bibinfo
  {author} {\bibfnamefont {A.}~\bibnamefont {Imamoglu}},\ and\ \bibinfo
  {author} {\bibfnamefont {M.~D.}\ \bibnamefont {Lukin}},\ }\bibfield  {title}
  {\bibinfo {title} {Laser cooling and real-time measurement of the nuclear
  spin environment of a solid-state qubit},\ }\href
  {https://doi.org/10.1038/nature10528} {\bibfield  {journal} {\bibinfo
  {journal} {Nature}\ }\textbf {\bibinfo {volume} {478}},\ \bibinfo {pages}
  {497} (\bibinfo {year} {2011})}\BibitemShut {NoStop}%
\bibitem [{\citenamefont {Yale}\ \emph {et~al.}(2013)\citenamefont {Yale},
  \citenamefont {Buckley}, \citenamefont {Christle}, \citenamefont {Burkard},
  \citenamefont {Heremans}, \citenamefont {Bassett},\ and\ \citenamefont
  {Awschalom}}]{yale13}%
  \BibitemOpen
  \bibfield  {author} {\bibinfo {author} {\bibfnamefont {C.~G.}\ \bibnamefont
  {Yale}}, \bibinfo {author} {\bibfnamefont {B.~B.}\ \bibnamefont {Buckley}},
  \bibinfo {author} {\bibfnamefont {D.~J.}\ \bibnamefont {Christle}}, \bibinfo
  {author} {\bibfnamefont {G.}~\bibnamefont {Burkard}}, \bibinfo {author}
  {\bibfnamefont {F.~J.}\ \bibnamefont {Heremans}}, \bibinfo {author}
  {\bibfnamefont {L.~C.}\ \bibnamefont {Bassett}},\ and\ \bibinfo {author}
  {\bibfnamefont {D.~D.}\ \bibnamefont {Awschalom}},\ }\bibfield  {title}
  {\bibinfo {title} {All-optical control of a solid-state spin using coherent
  dark states},\ }\href {https://doi.org/10.1073/pnas.1305920110} {\bibfield
  {journal} {\bibinfo  {journal} {Proceedings of the National Academy of
  Sciences}\ }\textbf {\bibinfo {volume} {110}},\ \bibinfo {pages} {7595}
  (\bibinfo {year} {2013})}\BibitemShut {NoStop}%
\bibitem [{\citenamefont {Golter}\ \emph {et~al.}(2013)\citenamefont {Golter},
  \citenamefont {Dinyari},\ and\ \citenamefont {Wang}}]{golter13}%
  \BibitemOpen
  \bibfield  {author} {\bibinfo {author} {\bibfnamefont {D.~A.}\ \bibnamefont
  {Golter}}, \bibinfo {author} {\bibfnamefont {K.~N.}\ \bibnamefont
  {Dinyari}},\ and\ \bibinfo {author} {\bibfnamefont {H.}~\bibnamefont
  {Wang}},\ }\bibfield  {title} {\bibinfo {title} {Nuclear-spin-dependent
  coherent population trapping of single nitrogen-vacancy centers in diamond},\
  }\href {https://doi.org/10.1103/physreva.87.035801} {\bibfield  {journal}
  {\bibinfo  {journal} {Phys. Rev. A}\ }\textbf {\bibinfo {volume} {87}},\
  \bibinfo {pages} {035801} (\bibinfo {year} {2013})}\BibitemShut {NoStop}%
\bibitem [{\citenamefont {Busaite}\ \emph {et~al.}(2020)\citenamefont
  {Busaite}, \citenamefont {Lazda}, \citenamefont {Berzins}, \citenamefont
  {Auzinsh}, \citenamefont {Ferber},\ and\ \citenamefont
  {Gahbauer}}]{busaite20}%
  \BibitemOpen
  \bibfield  {author} {\bibinfo {author} {\bibfnamefont {L.}~\bibnamefont
  {Busaite}}, \bibinfo {author} {\bibfnamefont {R.}~\bibnamefont {Lazda}},
  \bibinfo {author} {\bibfnamefont {A.}~\bibnamefont {Berzins}}, \bibinfo
  {author} {\bibfnamefont {M.}~\bibnamefont {Auzinsh}}, \bibinfo {author}
  {\bibfnamefont {R.}~\bibnamefont {Ferber}},\ and\ \bibinfo {author}
  {\bibfnamefont {F.}~\bibnamefont {Gahbauer}},\ }\bibfield  {title} {\bibinfo
  {title} {Dynamic {$^{14}$N} nuclear spin polarization in nitrogen-vacancy
  centers in diamond},\ }\href {https://doi.org/10.1103/physrevb.102.224101}
  {\bibfield  {journal} {\bibinfo  {journal} {Phys. Rev. B}\ }\textbf {\bibinfo
  {volume} {102}},\ \bibinfo {pages} {224101} (\bibinfo {year}
  {2020})}\BibitemShut {NoStop}%
\bibitem [{\citenamefont {Hegde}\ \emph {et~al.}(2020)\citenamefont {Hegde},
  \citenamefont {Zhang},\ and\ \citenamefont {Suter}}]{hegde20}%
  \BibitemOpen
  \bibfield  {author} {\bibinfo {author} {\bibfnamefont {S.~S.}\ \bibnamefont
  {Hegde}}, \bibinfo {author} {\bibfnamefont {J.}~\bibnamefont {Zhang}},\ and\
  \bibinfo {author} {\bibfnamefont {D.}~\bibnamefont {Suter}},\ }\bibfield
  {title} {\bibinfo {title} {Efficient quantum gates for individual nuclear
  spin qubits by indirect control},\ }\href
  {https://doi.org/10.1103/physrevlett.124.220501} {\bibfield  {journal}
  {\bibinfo  {journal} {Phys. Rev. Lett.}\ }\textbf {\bibinfo {volume} {124}},\
  \bibinfo {pages} {220501} (\bibinfo {year} {2020})}\BibitemShut {NoStop}%
\bibitem [{\citenamefont {Doherty}\ \emph {et~al.}(2013)\citenamefont
  {Doherty}, \citenamefont {Manson}, \citenamefont {Delaney}, \citenamefont
  {Jelezko}, \citenamefont {Wrachtrup},\ and\ \citenamefont
  {Hollenberg}}]{doherty13}%
  \BibitemOpen
  \bibfield  {author} {\bibinfo {author} {\bibfnamefont {M.~W.}\ \bibnamefont
  {Doherty}}, \bibinfo {author} {\bibfnamefont {N.~B.}\ \bibnamefont {Manson}},
  \bibinfo {author} {\bibfnamefont {P.}~\bibnamefont {Delaney}}, \bibinfo
  {author} {\bibfnamefont {F.}~\bibnamefont {Jelezko}}, \bibinfo {author}
  {\bibfnamefont {J.}~\bibnamefont {Wrachtrup}},\ and\ \bibinfo {author}
  {\bibfnamefont {L.~C.~L.}\ \bibnamefont {Hollenberg}},\ }\bibfield  {title}
  {\bibinfo {title} {{The nitrogen-vacancy colour centre in diamond}},\ }\href
  {https://doi.org/10.1016/j.physrep.2013.02.001} {\bibfield  {journal}
  {\bibinfo  {journal} {Phys. Rep.}\ }\textbf {\bibinfo {volume} {528}},\
  \bibinfo {pages} {1} (\bibinfo {year} {2013})}\BibitemShut {NoStop}%
\bibitem [{\citenamefont {Suter}\ and\ \citenamefont
  {Jelezko}(2017)}]{suter17}%
  \BibitemOpen
  \bibfield  {author} {\bibinfo {author} {\bibfnamefont {D.}~\bibnamefont
  {Suter}}\ and\ \bibinfo {author} {\bibfnamefont {F.}~\bibnamefont
  {Jelezko}},\ }\bibfield  {title} {\bibinfo {title} {Single-spin magnetic
  resonance in the nitrogen-vacancy center of diamond},\ }\href
  {https://doi.org/10.1016/j.pnmrs.2016.12.001} {\bibfield  {journal} {\bibinfo
   {journal} {Prog. Nucl. Magn. Reson. Spectrosc.}\ }\textbf {\bibinfo {volume}
  {98-99}},\ \bibinfo {pages} {50} (\bibinfo {year} {2017})}\BibitemShut
  {NoStop}%
\bibitem [{\citenamefont {Pezzagna}\ and\ \citenamefont
  {Meijer}(2021)}]{pezzagna21}%
  \BibitemOpen
  \bibfield  {author} {\bibinfo {author} {\bibfnamefont {S.}~\bibnamefont
  {Pezzagna}}\ and\ \bibinfo {author} {\bibfnamefont {J.}~\bibnamefont
  {Meijer}},\ }\bibfield  {title} {\bibinfo {title} {Quantum computer based on
  color centers in diamond},\ }\href {https://doi.org/10.1063/5.0007444}
  {\bibfield  {journal} {\bibinfo  {journal} {Applied Physics Reviews}\
  }\textbf {\bibinfo {volume} {8}},\ \bibinfo {pages} {011308} (\bibinfo {year}
  {2021})}\BibitemShut {NoStop}%
\bibitem [{\citenamefont {Thiering}\ and\ \citenamefont
  {Gali}(2018)}]{Thiering_2018}%
  \BibitemOpen
  \bibfield  {author} {\bibinfo {author} {\bibfnamefont {G.}~\bibnamefont
  {Thiering}}\ and\ \bibinfo {author} {\bibfnamefont {A.}~\bibnamefont
  {Gali}},\ }\bibfield  {title} {\bibinfo {title} {Theory of the optical
  spin-polarization loop of the nitrogen-vacancy center in diamond},\ }\href
  {https://doi.org/10.1103/PhysRevB.98.085207} {\bibfield  {journal} {\bibinfo
  {journal} {Phys. Rev. B}\ }\textbf {\bibinfo {volume} {98}},\ \bibinfo
  {pages} {085207} (\bibinfo {year} {2018})}\BibitemShut {NoStop}%
\bibitem [{\citenamefont {Gruber}\ \emph {et~al.}(1997)\citenamefont {Gruber},
  \citenamefont {Dräbenstedt}, \citenamefont {Tietz}, \citenamefont {Fleury},
  \citenamefont {Wrachtrup},\ and\ \citenamefont {von
  Borczyskowski}}]{Gruber_1997}%
  \BibitemOpen
  \bibfield  {author} {\bibinfo {author} {\bibfnamefont {A.}~\bibnamefont
  {Gruber}}, \bibinfo {author} {\bibfnamefont {A.}~\bibnamefont
  {Dräbenstedt}}, \bibinfo {author} {\bibfnamefont {C.}~\bibnamefont {Tietz}},
  \bibinfo {author} {\bibfnamefont {L.}~\bibnamefont {Fleury}}, \bibinfo
  {author} {\bibfnamefont {J.}~\bibnamefont {Wrachtrup}},\ and\ \bibinfo
  {author} {\bibfnamefont {C.}~\bibnamefont {von Borczyskowski}},\ }\bibfield
  {title} {\bibinfo {title} {Scanning confocal optical microscopy and magnetic
  resonance on single defect centers},\ }\href
  {https://doi.org/10.1126/science.276.5321.2012} {\bibfield  {journal}
  {\bibinfo  {journal} {Science}\ }\textbf {\bibinfo {volume} {276}},\ \bibinfo
  {pages} {2012} (\bibinfo {year} {1997})},\ \Eprint
  {https://arxiv.org/abs/https://www.science.org/doi/pdf/10.1126/science.276.5321.2012}
  {https://www.science.org/doi/pdf/10.1126/science.276.5321.2012} \BibitemShut
  {NoStop}%
\bibitem [{\citenamefont {Wellmann}(2018)}]{Wellmann_2018}%
  \BibitemOpen
  \bibfield  {author} {\bibinfo {author} {\bibfnamefont {P.~J.}\ \bibnamefont
  {Wellmann}},\ }\bibfield  {title} {\bibinfo {title} {Review of {S}i{C}
  crystal growth technology},\ }\href
  {https://doi.org/10.1088/1361-6641/aad831} {\bibfield  {journal} {\bibinfo
  {journal} {Semiconductor Science and Technology}\ }\textbf {\bibinfo {volume}
  {33}},\ \bibinfo {pages} {103001} (\bibinfo {year} {2018})}\BibitemShut
  {NoStop}%
\bibitem [{\citenamefont {Liu}\ \emph {et~al.}(2020)\citenamefont {Liu},
  \citenamefont {Xu}, \citenamefont {Song}, \citenamefont {Wang}, \citenamefont
  {Dong}, \citenamefont {Li}, \citenamefont {Ren}, \citenamefont {Li},
  \citenamefont {Rommel}, \citenamefont {Gu}, \citenamefont {Liu},
  \citenamefont {Hu},\ and\ \citenamefont {Fang}}]{Liu_2020}%
  \BibitemOpen
  \bibfield  {author} {\bibinfo {author} {\bibfnamefont {J.}~\bibnamefont
  {Liu}}, \bibinfo {author} {\bibfnamefont {Z.}~\bibnamefont {Xu}}, \bibinfo
  {author} {\bibfnamefont {Y.}~\bibnamefont {Song}}, \bibinfo {author}
  {\bibfnamefont {H.}~\bibnamefont {Wang}}, \bibinfo {author} {\bibfnamefont
  {B.}~\bibnamefont {Dong}}, \bibinfo {author} {\bibfnamefont {S.}~\bibnamefont
  {Li}}, \bibinfo {author} {\bibfnamefont {J.}~\bibnamefont {Ren}}, \bibinfo
  {author} {\bibfnamefont {Q.}~\bibnamefont {Li}}, \bibinfo {author}
  {\bibfnamefont {M.}~\bibnamefont {Rommel}}, \bibinfo {author} {\bibfnamefont
  {X.}~\bibnamefont {Gu}}, \bibinfo {author} {\bibfnamefont {B.}~\bibnamefont
  {Liu}}, \bibinfo {author} {\bibfnamefont {M.}~\bibnamefont {Hu}},\ and\
  \bibinfo {author} {\bibfnamefont {F.}~\bibnamefont {Fang}},\ }\bibfield
  {title} {\bibinfo {title} {{Confocal photoluminescence characterization of
  silicon-vacancy color centers in 4H-SiC fabricated by a femtosecond laser}},\
  }\href {https://doi.org/10.1016/j.npe.2020.11.003} {\bibfield  {journal}
  {\bibinfo  {journal} {Nanotechnology and Precision Engineering (NPE)}\
  }\textbf {\bibinfo {volume} {3}},\ \bibinfo {pages} {218} (\bibinfo {year}
  {2020})},\ \Eprint
  {https://arxiv.org/abs/https://pubs.aip.org/tu/npe/article-pdf/3/4/218/16662950/218\_1\_online.pdf}
  {https://pubs.aip.org/tu/npe/article-pdf/3/4/218/16662950/218\_1\_online.pdf}
  \BibitemShut {NoStop}%
\bibitem [{\citenamefont {Chakravorty}\ \emph {et~al.}(2021)\citenamefont
  {Chakravorty}, \citenamefont {Singh}, \citenamefont {Jatav}, \citenamefont
  {Meena}, \citenamefont {Kanjilal},\ and\ \citenamefont
  {Kabiraj}}]{Chakravorty_2021}%
  \BibitemOpen
  \bibfield  {author} {\bibinfo {author} {\bibfnamefont {A.}~\bibnamefont
  {Chakravorty}}, \bibinfo {author} {\bibfnamefont {B.}~\bibnamefont {Singh}},
  \bibinfo {author} {\bibfnamefont {H.}~\bibnamefont {Jatav}}, \bibinfo
  {author} {\bibfnamefont {R.}~\bibnamefont {Meena}}, \bibinfo {author}
  {\bibfnamefont {D.}~\bibnamefont {Kanjilal}},\ and\ \bibinfo {author}
  {\bibfnamefont {D.}~\bibnamefont {Kabiraj}},\ }\bibfield  {title} {\bibinfo
  {title} {{Controlled generation of photoemissive defects in 4H-SiC using
  swift heavy ion irradiation}},\ }\href {https://doi.org/10.1063/5.0051328}
  {\bibfield  {journal} {\bibinfo  {journal} {Journal of Applied Physics}\
  }\textbf {\bibinfo {volume} {129}},\ \bibinfo {pages} {245905} (\bibinfo
  {year} {2021})},\ \Eprint
  {https://arxiv.org/abs/https://pubs.aip.org/aip/jap/article-pdf/doi/10.1063/5.0051328/15268515/245905\_1\_online.pdf}
  {https://pubs.aip.org/aip/jap/article-pdf/doi/10.1063/5.0051328/15268515/245905\_1\_online.pdf}
  \BibitemShut {NoStop}%
\bibitem [{\citenamefont {Song}\ \emph {et~al.}(2019)\citenamefont {Song},
  \citenamefont {Asano}, \citenamefont {Jeon}, \citenamefont {Kim},
  \citenamefont {Chen}, \citenamefont {Kang},\ and\ \citenamefont
  {Noda}}]{Song_2019}%
  \BibitemOpen
  \bibfield  {author} {\bibinfo {author} {\bibfnamefont {B.-S.}\ \bibnamefont
  {Song}}, \bibinfo {author} {\bibfnamefont {T.}~\bibnamefont {Asano}},
  \bibinfo {author} {\bibfnamefont {S.}~\bibnamefont {Jeon}}, \bibinfo {author}
  {\bibfnamefont {H.}~\bibnamefont {Kim}}, \bibinfo {author} {\bibfnamefont
  {C.}~\bibnamefont {Chen}}, \bibinfo {author} {\bibfnamefont {D.~D.}\
  \bibnamefont {Kang}},\ and\ \bibinfo {author} {\bibfnamefont
  {S.}~\bibnamefont {Noda}},\ }\bibfield  {title} {\bibinfo {title}
  {Ultrahigh-{Q} photonic crystal nanocavities based on 4{H} silicon carbide},\
  }\href {https://doi.org/10.1364/OPTICA.6.000991} {\bibfield  {journal}
  {\bibinfo  {journal} {Optica}\ }\textbf {\bibinfo {volume} {6}},\ \bibinfo
  {pages} {991} (\bibinfo {year} {2019})}\BibitemShut {NoStop}%
\bibitem [{\citenamefont {Lukin}\ \emph {et~al.}(2020)\citenamefont {Lukin},
  \citenamefont {Dory}, \citenamefont {Guidry}, \citenamefont {Yang},
  \citenamefont {Mishra}, \citenamefont {Trivedi}, \citenamefont {Radulaski},
  \citenamefont {Sun}, \citenamefont {Vercruysse}, \citenamefont {Ahn},\ and\
  \citenamefont {Vu{\v{c}}kovi{\'{c}}}}]{Lukin_2020}%
  \BibitemOpen
  \bibfield  {author} {\bibinfo {author} {\bibfnamefont {D.~M.}\ \bibnamefont
  {Lukin}}, \bibinfo {author} {\bibfnamefont {C.}~\bibnamefont {Dory}},
  \bibinfo {author} {\bibfnamefont {M.~A.}\ \bibnamefont {Guidry}}, \bibinfo
  {author} {\bibfnamefont {K.~Y.}\ \bibnamefont {Yang}}, \bibinfo {author}
  {\bibfnamefont {S.~D.}\ \bibnamefont {Mishra}}, \bibinfo {author}
  {\bibfnamefont {R.}~\bibnamefont {Trivedi}}, \bibinfo {author} {\bibfnamefont
  {M.}~\bibnamefont {Radulaski}}, \bibinfo {author} {\bibfnamefont
  {S.}~\bibnamefont {Sun}}, \bibinfo {author} {\bibfnamefont {D.}~\bibnamefont
  {Vercruysse}}, \bibinfo {author} {\bibfnamefont {G.~H.}\ \bibnamefont
  {Ahn}},\ and\ \bibinfo {author} {\bibfnamefont {J.}~\bibnamefont
  {Vu{\v{c}}kovi{\'{c}}}},\ }\bibfield  {title} {\bibinfo {title}
  {4{H}-silicon-carbide-on-insulator for integrated quantum and nonlinear
  photonics},\ }\href {https://doi.org/10.1038/s41566-019-0556-6} {\bibfield
  {journal} {\bibinfo  {journal} {Nature Photonics}\ }\textbf {\bibinfo
  {volume} {14}},\ \bibinfo {pages} {330} (\bibinfo {year} {2020})}\BibitemShut
  {NoStop}%
\bibitem [{\citenamefont {Guidry}\ \emph {et~al.}(2020)\citenamefont {Guidry},
  \citenamefont {Yang}, \citenamefont {Lukin}, \citenamefont {Markosyan},
  \citenamefont {Yang}, \citenamefont {Fejer},\ and\ \citenamefont
  {Vu\v{c}kovi\'{c}}}]{Guidry_2020}%
  \BibitemOpen
  \bibfield  {author} {\bibinfo {author} {\bibfnamefont {M.~A.}\ \bibnamefont
  {Guidry}}, \bibinfo {author} {\bibfnamefont {K.~Y.}\ \bibnamefont {Yang}},
  \bibinfo {author} {\bibfnamefont {D.~M.}\ \bibnamefont {Lukin}}, \bibinfo
  {author} {\bibfnamefont {A.}~\bibnamefont {Markosyan}}, \bibinfo {author}
  {\bibfnamefont {J.}~\bibnamefont {Yang}}, \bibinfo {author} {\bibfnamefont
  {M.~M.}\ \bibnamefont {Fejer}},\ and\ \bibinfo {author} {\bibfnamefont
  {J.}~\bibnamefont {Vu\v{c}kovi\'{c}}},\ }\bibfield  {title} {\bibinfo {title}
  {Optical parametric oscillation in silicon carbide nanophotonics},\ }\href
  {https://doi.org/10.1364/OPTICA.394138} {\bibfield  {journal} {\bibinfo
  {journal} {Optica}\ }\textbf {\bibinfo {volume} {7}},\ \bibinfo {pages}
  {1139} (\bibinfo {year} {2020})}\BibitemShut {NoStop}%
\bibitem [{\citenamefont {Radulaski}\ \emph {et~al.}(2017)\citenamefont
  {Radulaski}, \citenamefont {Widmann}, \citenamefont {Niethammer},
  \citenamefont {Zhang}, \citenamefont {Lee}, \citenamefont {Rendler},
  \citenamefont {Lagoudakis}, \citenamefont {Son}, \citenamefont {Janzén},
  \citenamefont {Ohshima}, \citenamefont {Wrachtrup},\ and\ \citenamefont
  {Vučković}}]{Radulaski_2017}%
  \BibitemOpen
  \bibfield  {author} {\bibinfo {author} {\bibfnamefont {M.}~\bibnamefont
  {Radulaski}}, \bibinfo {author} {\bibfnamefont {M.}~\bibnamefont {Widmann}},
  \bibinfo {author} {\bibfnamefont {M.}~\bibnamefont {Niethammer}}, \bibinfo
  {author} {\bibfnamefont {J.~L.}\ \bibnamefont {Zhang}}, \bibinfo {author}
  {\bibfnamefont {S.-Y.}\ \bibnamefont {Lee}}, \bibinfo {author} {\bibfnamefont
  {T.}~\bibnamefont {Rendler}}, \bibinfo {author} {\bibfnamefont {K.~G.}\
  \bibnamefont {Lagoudakis}}, \bibinfo {author} {\bibfnamefont {N.~T.}\
  \bibnamefont {Son}}, \bibinfo {author} {\bibfnamefont {E.}~\bibnamefont
  {Janzén}}, \bibinfo {author} {\bibfnamefont {T.}~\bibnamefont {Ohshima}},
  \bibinfo {author} {\bibfnamefont {J.}~\bibnamefont {Wrachtrup}},\ and\
  \bibinfo {author} {\bibfnamefont {J.}~\bibnamefont {Vučković}},\ }\bibfield
   {title} {\bibinfo {title} {Scalable quantum photonics with single color
  centers in silicon carbide},\ }\href
  {https://doi.org/10.1021/acs.nanolett.6b05102} {\bibfield  {journal}
  {\bibinfo  {journal} {Nano Letters}\ }\textbf {\bibinfo {volume} {17}},\
  \bibinfo {pages} {1782} (\bibinfo {year} {2017})},\ \bibinfo {note} {pMID:
  28225630},\ \Eprint
  {https://arxiv.org/abs/https://doi.org/10.1021/acs.nanolett.6b05102}
  {https://doi.org/10.1021/acs.nanolett.6b05102} \BibitemShut {NoStop}%
\bibitem [{\citenamefont {Wang}\ \emph {et~al.}(2017)\citenamefont {Wang},
  \citenamefont {Zhou}, \citenamefont {Zhang}, \citenamefont {Liu},
  \citenamefont {Li}, \citenamefont {Li}, \citenamefont {Liu}, \citenamefont
  {Wang},\ and\ \citenamefont {Gao}}]{Wang_2017}%
  \BibitemOpen
  \bibfield  {author} {\bibinfo {author} {\bibfnamefont {J.}~\bibnamefont
  {Wang}}, \bibinfo {author} {\bibfnamefont {Y.}~\bibnamefont {Zhou}}, \bibinfo
  {author} {\bibfnamefont {X.}~\bibnamefont {Zhang}}, \bibinfo {author}
  {\bibfnamefont {F.}~\bibnamefont {Liu}}, \bibinfo {author} {\bibfnamefont
  {Y.}~\bibnamefont {Li}}, \bibinfo {author} {\bibfnamefont {K.}~\bibnamefont
  {Li}}, \bibinfo {author} {\bibfnamefont {Z.}~\bibnamefont {Liu}}, \bibinfo
  {author} {\bibfnamefont {G.}~\bibnamefont {Wang}},\ and\ \bibinfo {author}
  {\bibfnamefont {W.}~\bibnamefont {Gao}},\ }\bibfield  {title} {\bibinfo
  {title} {Efficient generation of an array of single silicon-vacancy defects
  in silicon carbide},\ }\href
  {https://doi.org/10.1103/PhysRevApplied.7.064021} {\bibfield  {journal}
  {\bibinfo  {journal} {Phys. Rev. Appl.}\ }\textbf {\bibinfo {volume} {7}},\
  \bibinfo {pages} {064021} (\bibinfo {year} {2017})}\BibitemShut {NoStop}%
\bibitem [{\citenamefont {S\"orman}\ \emph {et~al.}(2000)\citenamefont
  {S\"orman}, \citenamefont {Son}, \citenamefont {Chen}, \citenamefont
  {Kordina}, \citenamefont {Hallin},\ and\ \citenamefont
  {Janz\'en}}]{Sorman_2000}%
  \BibitemOpen
  \bibfield  {author} {\bibinfo {author} {\bibfnamefont {E.}~\bibnamefont
  {S\"orman}}, \bibinfo {author} {\bibfnamefont {N.~T.}\ \bibnamefont {Son}},
  \bibinfo {author} {\bibfnamefont {W.~M.}\ \bibnamefont {Chen}}, \bibinfo
  {author} {\bibfnamefont {O.}~\bibnamefont {Kordina}}, \bibinfo {author}
  {\bibfnamefont {C.}~\bibnamefont {Hallin}},\ and\ \bibinfo {author}
  {\bibfnamefont {E.}~\bibnamefont {Janz\'en}},\ }\bibfield  {title} {\bibinfo
  {title} {Silicon vacancy related defect in 4{H} and 6{H} {S}i{C}},\ }\href
  {https://doi.org/10.1103/PhysRevB.61.2613} {\bibfield  {journal} {\bibinfo
  {journal} {Phys. Rev. B}\ }\textbf {\bibinfo {volume} {61}},\ \bibinfo
  {pages} {2613} (\bibinfo {year} {2000})}\BibitemShut {NoStop}%
\bibitem [{\citenamefont {Janzén}\ \emph {et~al.}(2009)\citenamefont
  {Janzén}, \citenamefont {Gali}, \citenamefont {Carlsson}, \citenamefont
  {Gällström}, \citenamefont {Magnusson},\ and\ \citenamefont
  {Son}}]{Janzen_2009}%
  \BibitemOpen
  \bibfield  {author} {\bibinfo {author} {\bibfnamefont {E.}~\bibnamefont
  {Janzén}}, \bibinfo {author} {\bibfnamefont {A.}~\bibnamefont {Gali}},
  \bibinfo {author} {\bibfnamefont {P.}~\bibnamefont {Carlsson}}, \bibinfo
  {author} {\bibfnamefont {A.}~\bibnamefont {Gällström}}, \bibinfo {author}
  {\bibfnamefont {B.}~\bibnamefont {Magnusson}},\ and\ \bibinfo {author}
  {\bibfnamefont {N.}~\bibnamefont {Son}},\ }\bibfield  {title} {\bibinfo
  {title} {The silicon vacancy in {S}i{C}},\ }\href
  {https://doi.org/https://doi.org/10.1016/j.physb.2009.09.023} {\bibfield
  {journal} {\bibinfo  {journal} {Physica B: Condensed Matter}\ }\textbf
  {\bibinfo {volume} {404}},\ \bibinfo {pages} {4354} (\bibinfo {year}
  {2009})}\BibitemShut {NoStop}%
\bibitem [{\citenamefont {Nagy}\ \emph {et~al.}(2018)\citenamefont {Nagy},
  \citenamefont {Widmann}, \citenamefont {Niethammer}, \citenamefont {Dasari},
  \citenamefont {Gerhardt}, \citenamefont {Soykal}, \citenamefont {Radulaski},
  \citenamefont {Ohshima}, \citenamefont {Vu\ifmmode \check{c}\else
  \v{c}\fi{}kovi\ifmmode~\acute{c}\else \'{c}\fi{}}, \citenamefont {Son},
  \citenamefont {Ivanov}, \citenamefont {Economou}, \citenamefont {Bonato},
  \citenamefont {Lee},\ and\ \citenamefont {Wrachtrup}}]{Nagy_2018}%
  \BibitemOpen
  \bibfield  {author} {\bibinfo {author} {\bibfnamefont {R.}~\bibnamefont
  {Nagy}}, \bibinfo {author} {\bibfnamefont {M.}~\bibnamefont {Widmann}},
  \bibinfo {author} {\bibfnamefont {M.}~\bibnamefont {Niethammer}}, \bibinfo
  {author} {\bibfnamefont {D.~B.~R.}\ \bibnamefont {Dasari}}, \bibinfo {author}
  {\bibfnamefont {I.}~\bibnamefont {Gerhardt}}, \bibinfo {author}
  {\bibfnamefont {O.~O.}\ \bibnamefont {Soykal}}, \bibinfo {author}
  {\bibfnamefont {M.}~\bibnamefont {Radulaski}}, \bibinfo {author}
  {\bibfnamefont {T.}~\bibnamefont {Ohshima}}, \bibinfo {author} {\bibfnamefont
  {J.}~\bibnamefont {Vu\ifmmode \check{c}\else
  \v{c}\fi{}kovi\ifmmode~\acute{c}\else \'{c}\fi{}}}, \bibinfo {author}
  {\bibfnamefont {N.~T.}\ \bibnamefont {Son}}, \bibinfo {author} {\bibfnamefont
  {I.~G.}\ \bibnamefont {Ivanov}}, \bibinfo {author} {\bibfnamefont {S.~E.}\
  \bibnamefont {Economou}}, \bibinfo {author} {\bibfnamefont {C.}~\bibnamefont
  {Bonato}}, \bibinfo {author} {\bibfnamefont {S.-Y.}\ \bibnamefont {Lee}},\
  and\ \bibinfo {author} {\bibfnamefont {J.}~\bibnamefont {Wrachtrup}},\
  }\bibfield  {title} {\bibinfo {title} {Quantum properties of dichroic silicon
  vacancies in silicon carbide},\ }\href
  {https://doi.org/10.1103/PhysRevApplied.9.034022} {\bibfield  {journal}
  {\bibinfo  {journal} {Phys. Rev. Appl.}\ }\textbf {\bibinfo {volume} {9}},\
  \bibinfo {pages} {034022} (\bibinfo {year} {2018})}\BibitemShut {NoStop}%
\bibitem [{\citenamefont {Gali}\ \emph {et~al.}(2010)\citenamefont {Gali},
  \citenamefont {G{\"{a}}llstr{\"{o}}m}, \citenamefont {Son},\ and\
  \citenamefont {Janz{\'{e}}n}}]{Gali_2010}%
  \BibitemOpen
  \bibfield  {author} {\bibinfo {author} {\bibfnamefont {A.}~\bibnamefont
  {Gali}}, \bibinfo {author} {\bibfnamefont {A.}~\bibnamefont
  {G{\"{a}}llstr{\"{o}}m}}, \bibinfo {author} {\bibfnamefont {N.~T.}\
  \bibnamefont {Son}},\ and\ \bibinfo {author} {\bibfnamefont {E.}~\bibnamefont
  {Janz{\'{e}}n}},\ }\bibfield  {title} {\bibinfo {title} {Theory of neutral
  divacancy in sic: A defect for spintronics},\ }in\ \href
  {https://doi.org/10.4028/www.scientific.net/MSF.645-648.395} {\emph {\bibinfo
  {booktitle} {Silicon Carbide and Related Materials 2009}}},\ \bibinfo
  {series} {Materials Science Forum}, Vol.\ \bibinfo {volume} {645}\ (\bibinfo
  {publisher} {Trans Tech Publications Ltd},\ \bibinfo {year} {2010})\ pp.\
  \bibinfo {pages} {395--397}\BibitemShut {NoStop}%
\bibitem [{\citenamefont {Falk}\ \emph {et~al.}(2013)\citenamefont {Falk},
  \citenamefont {Buckley}, \citenamefont {Calusine}, \citenamefont {Koehl},
  \citenamefont {Dobrovitski}, \citenamefont {Politi}, \citenamefont {Zorman},
  \citenamefont {Feng},\ and\ \citenamefont {Awschalom}}]{Falk_2013}%
  \BibitemOpen
  \bibfield  {author} {\bibinfo {author} {\bibfnamefont {A.~L.}\ \bibnamefont
  {Falk}}, \bibinfo {author} {\bibfnamefont {B.~B.}\ \bibnamefont {Buckley}},
  \bibinfo {author} {\bibfnamefont {G.}~\bibnamefont {Calusine}}, \bibinfo
  {author} {\bibfnamefont {W.~F.}\ \bibnamefont {Koehl}}, \bibinfo {author}
  {\bibfnamefont {V.~V.}\ \bibnamefont {Dobrovitski}}, \bibinfo {author}
  {\bibfnamefont {A.}~\bibnamefont {Politi}}, \bibinfo {author} {\bibfnamefont
  {C.~A.}\ \bibnamefont {Zorman}}, \bibinfo {author} {\bibfnamefont {P.~X.-L.}\
  \bibnamefont {Feng}},\ and\ \bibinfo {author} {\bibfnamefont {D.~D.}\
  \bibnamefont {Awschalom}},\ }\bibfield  {title} {\bibinfo {title} {Polytype
  control of spin qubits in silicon carbide},\ }\href
  {https://doi.org/10.1038/ncomms2854} {\bibfield  {journal} {\bibinfo
  {journal} {Nature Communications}\ }\textbf {\bibinfo {volume} {4}},\
  \bibinfo {pages} {1819} (\bibinfo {year} {2013})}\BibitemShut {NoStop}%
\bibitem [{\citenamefont {Bosma}\ \emph {et~al.}(2018)\citenamefont {Bosma},
  \citenamefont {Lof}, \citenamefont {Gilardoni}, \citenamefont {Zwier},
  \citenamefont {Hendriks}, \citenamefont {Magnusson}, \citenamefont {Ellison},
  \citenamefont {G\"allstr\"om}, \citenamefont {Ivanov}, \citenamefont {Son},
  \citenamefont {Havenith},\ and\ \citenamefont {van~der Wal}}]{bosma18}%
  \BibitemOpen
  \bibfield  {author} {\bibinfo {author} {\bibfnamefont {T.}~\bibnamefont
  {Bosma}}, \bibinfo {author} {\bibfnamefont {G.~J.~J.}\ \bibnamefont {Lof}},
  \bibinfo {author} {\bibfnamefont {C.~M.}\ \bibnamefont {Gilardoni}}, \bibinfo
  {author} {\bibfnamefont {O.~V.}\ \bibnamefont {Zwier}}, \bibinfo {author}
  {\bibfnamefont {F.}~\bibnamefont {Hendriks}}, \bibinfo {author}
  {\bibfnamefont {B.}~\bibnamefont {Magnusson}}, \bibinfo {author}
  {\bibfnamefont {A.}~\bibnamefont {Ellison}}, \bibinfo {author} {\bibfnamefont
  {A.}~\bibnamefont {G\"allstr\"om}}, \bibinfo {author} {\bibfnamefont {I.~G.}\
  \bibnamefont {Ivanov}}, \bibinfo {author} {\bibfnamefont {N.~T.}\
  \bibnamefont {Son}}, \bibinfo {author} {\bibfnamefont {R.~W.~A.}\
  \bibnamefont {Havenith}},\ and\ \bibinfo {author} {\bibfnamefont {C.~H.}\
  \bibnamefont {van~der Wal}},\ }\bibfield  {title} {\bibinfo {title}
  {Identification and tunable optical coherent control of transition-metal
  spins in silicon carbide},\ }\href
  {https://doi.org/10.1038/s41534-018-0097-8} {\bibfield  {journal} {\bibinfo
  {journal} {npj Quantum Inf.}\ }\textbf {\bibinfo {volume} {4}},\ \bibinfo
  {pages} {48} (\bibinfo {year} {2018})}\BibitemShut {NoStop}%
\bibitem [{\citenamefont {Spindlberger}\ \emph {et~al.}(2019)\citenamefont
  {Spindlberger}, \citenamefont {Cs\'or\'e}, \citenamefont {Thiering},
  \citenamefont {Putz}, \citenamefont {Karhu}, \citenamefont {Hassan},
  \citenamefont {Son}, \citenamefont {Fromherz}, \citenamefont {Gali},\ and\
  \citenamefont {Trupke}}]{spindlberger19}%
  \BibitemOpen
  \bibfield  {author} {\bibinfo {author} {\bibfnamefont {L.}~\bibnamefont
  {Spindlberger}}, \bibinfo {author} {\bibfnamefont {A.}~\bibnamefont
  {Cs\'or\'e}}, \bibinfo {author} {\bibfnamefont {G.}~\bibnamefont {Thiering}},
  \bibinfo {author} {\bibfnamefont {S.}~\bibnamefont {Putz}}, \bibinfo {author}
  {\bibfnamefont {R.}~\bibnamefont {Karhu}}, \bibinfo {author} {\bibfnamefont
  {J.}~\bibnamefont {Hassan}}, \bibinfo {author} {\bibfnamefont
  {N.}~\bibnamefont {Son}}, \bibinfo {author} {\bibfnamefont {T.}~\bibnamefont
  {Fromherz}}, \bibinfo {author} {\bibfnamefont {A.}~\bibnamefont {Gali}},\
  and\ \bibinfo {author} {\bibfnamefont {M.}~\bibnamefont {Trupke}},\
  }\bibfield  {title} {\bibinfo {title} {Optical properties of vanadium in 4{H}
  silicon carbide for quantum technology},\ }\href
  {https://doi.org/10.1103/physrevapplied.12.014015} {\bibfield  {journal}
  {\bibinfo  {journal} {Phys. Rev. Appl.}\ }\textbf {\bibinfo {volume} {12}},\
  \bibinfo {pages} {014015} (\bibinfo {year} {2019})}\BibitemShut {NoStop}%
\bibitem [{\citenamefont {Gilardoni}\ \emph {et~al.}(2020)\citenamefont
  {Gilardoni}, \citenamefont {Bosma}, \citenamefont {van Hien}, \citenamefont
  {Hendriks}, \citenamefont {Magnusson}, \citenamefont {Ellison}, \citenamefont
  {Ivanov}, \citenamefont {Son},\ and\ \citenamefont {van~der
  Wal}}]{gilardoni20}%
  \BibitemOpen
  \bibfield  {author} {\bibinfo {author} {\bibfnamefont {C.~M.}\ \bibnamefont
  {Gilardoni}}, \bibinfo {author} {\bibfnamefont {T.}~\bibnamefont {Bosma}},
  \bibinfo {author} {\bibfnamefont {D.}~\bibnamefont {van Hien}}, \bibinfo
  {author} {\bibfnamefont {F.}~\bibnamefont {Hendriks}}, \bibinfo {author}
  {\bibfnamefont {B.}~\bibnamefont {Magnusson}}, \bibinfo {author}
  {\bibfnamefont {A.}~\bibnamefont {Ellison}}, \bibinfo {author} {\bibfnamefont
  {I.~G.}\ \bibnamefont {Ivanov}}, \bibinfo {author} {\bibfnamefont {N.~T.}\
  \bibnamefont {Son}},\ and\ \bibinfo {author} {\bibfnamefont {C.~H.}\
  \bibnamefont {van~der Wal}},\ }\bibfield  {title} {\bibinfo {title}
  {Spin-relaxation times exceeding seconds for color centers with strong
  spin–orbit coupling in {SiC}},\ }\href
  {https://doi.org/10.1088/1367-2630/abbf23} {\bibfield  {journal} {\bibinfo
  {journal} {New J. Phys.}\ }\textbf {\bibinfo {volume} {22}},\ \bibinfo
  {pages} {103051} (\bibinfo {year} {2020})}\BibitemShut {NoStop}%
\bibitem [{\citenamefont {Wolfowicz}\ \emph {et~al.}(2020)\citenamefont
  {Wolfowicz}, \citenamefont {Anderson}, \citenamefont {Diler}, \citenamefont
  {Poluektov}, \citenamefont {Heremans},\ and\ \citenamefont
  {Awschalom}}]{wolfowicz20}%
  \BibitemOpen
  \bibfield  {author} {\bibinfo {author} {\bibfnamefont {G.}~\bibnamefont
  {Wolfowicz}}, \bibinfo {author} {\bibfnamefont {C.~P.}\ \bibnamefont
  {Anderson}}, \bibinfo {author} {\bibfnamefont {B.}~\bibnamefont {Diler}},
  \bibinfo {author} {\bibfnamefont {O.~G.}\ \bibnamefont {Poluektov}}, \bibinfo
  {author} {\bibfnamefont {F.~J.}\ \bibnamefont {Heremans}},\ and\ \bibinfo
  {author} {\bibfnamefont {D.~D.}\ \bibnamefont {Awschalom}},\ }\bibfield
  {title} {\bibinfo {title} {Vanadium spin qubits as telecom quantum emitters
  in silicon carbide},\ }\href {https://doi.org/10.1126/sciadv.aaz1192}
  {\bibfield  {journal} {\bibinfo  {journal} {Sci. Adv.}\ }\textbf {\bibinfo
  {volume} {6}},\ \bibinfo {pages} {eaaz1192} (\bibinfo {year}
  {2020})}\BibitemShut {NoStop}%
\bibitem [{\citenamefont {Astner}\ \emph {et~al.}(2022)\citenamefont {Astner},
  \citenamefont {Koller}, \citenamefont {Gilardoni}, \citenamefont {Hendriks},
  \citenamefont {Son}, \citenamefont {Ivanov}, \citenamefont {Hassan},
  \citenamefont {Wal},\ and\ \citenamefont {Trupke}}]{astner22}%
  \BibitemOpen
  \bibfield  {author} {\bibinfo {author} {\bibfnamefont {T.}~\bibnamefont
  {Astner}}, \bibinfo {author} {\bibfnamefont {P.}~\bibnamefont {Koller}},
  \bibinfo {author} {\bibfnamefont {C.~M.}\ \bibnamefont {Gilardoni}}, \bibinfo
  {author} {\bibfnamefont {J.}~\bibnamefont {Hendriks}}, \bibinfo {author}
  {\bibfnamefont {N.~T.}\ \bibnamefont {Son}}, \bibinfo {author} {\bibfnamefont
  {I.~G.}\ \bibnamefont {Ivanov}}, \bibinfo {author} {\bibfnamefont {J.~U.}\
  \bibnamefont {Hassan}}, \bibinfo {author} {\bibfnamefont {C.~H. v.~d.}\
  \bibnamefont {Wal}},\ and\ \bibinfo {author} {\bibfnamefont {M.}~\bibnamefont
  {Trupke}},\ }\bibfield  {title} {\bibinfo {title} {Vanadium in silicon
  carbide: Telecom-ready spin centres with long relaxation lifetimes and
  hyperfine-resolved optical transitions},\ }\href
  {http://arxiv.org/abs/2206.06240v1} {\bibfield  {journal} {\bibinfo
  {journal} {arXiv}\ } (\bibinfo {year} {2022})},\ \Eprint
  {https://arxiv.org/abs/2206.06240} {arXiv:2206.06240 [quant-ph]} \BibitemShut
  {NoStop}%
\bibitem [{\citenamefont {Cilibrizzi}\ \emph {et~al.}(2023)\citenamefont
  {Cilibrizzi}, \citenamefont {Arshad}, \citenamefont {Tissot}, \citenamefont
  {Son}, \citenamefont {Ivanov}, \citenamefont {Astner}, \citenamefont
  {Koller}, \citenamefont {Ghezellou}, \citenamefont {Ul-Hassan}, \citenamefont
  {White}, \citenamefont {Bekker}, \citenamefont {Burkard}, \citenamefont
  {Trupke},\ and\ \citenamefont {Bonato}}]{cilibrizzi23}%
  \BibitemOpen
  \bibfield  {author} {\bibinfo {author} {\bibfnamefont {P.}~\bibnamefont
  {Cilibrizzi}}, \bibinfo {author} {\bibfnamefont {M.~J.}\ \bibnamefont
  {Arshad}}, \bibinfo {author} {\bibfnamefont {B.}~\bibnamefont {Tissot}},
  \bibinfo {author} {\bibfnamefont {N.~T.}\ \bibnamefont {Son}}, \bibinfo
  {author} {\bibfnamefont {I.~G.}\ \bibnamefont {Ivanov}}, \bibinfo {author}
  {\bibfnamefont {T.}~\bibnamefont {Astner}}, \bibinfo {author} {\bibfnamefont
  {P.}~\bibnamefont {Koller}}, \bibinfo {author} {\bibfnamefont
  {M.}~\bibnamefont {Ghezellou}}, \bibinfo {author} {\bibfnamefont
  {J.}~\bibnamefont {Ul-Hassan}}, \bibinfo {author} {\bibfnamefont
  {D.}~\bibnamefont {White}}, \bibinfo {author} {\bibfnamefont
  {C.}~\bibnamefont {Bekker}}, \bibinfo {author} {\bibfnamefont
  {G.}~\bibnamefont {Burkard}}, \bibinfo {author} {\bibfnamefont
  {M.}~\bibnamefont {Trupke}},\ and\ \bibinfo {author} {\bibfnamefont
  {C.}~\bibnamefont {Bonato}},\ }\bibfield  {title} {\bibinfo {title}
  {Ultra-narrow inhomogeneous spectral distribution of telecom-wavelength
  vanadium centres in isotopically-enriched silicon carbide},\ }\bibfield
  {journal} {\bibinfo  {journal} {arXiv}\ }\href
  {https://doi.org/10.48550/ARXIV.2305.01757} {10.48550/ARXIV.2305.01757}
  (\bibinfo {year} {2023}),\ \Eprint {https://arxiv.org/abs/2206.06240}
  {arXiv:2206.06240 [quant-ph]} \BibitemShut {NoStop}%
\bibitem [{\citenamefont {Cs{\'o}r{\'e}}\ and\ \citenamefont
  {Gali}(2020)}]{csore20}%
  \BibitemOpen
  \bibfield  {author} {\bibinfo {author} {\bibfnamefont {A.}~\bibnamefont
  {Cs{\'o}r{\'e}}}\ and\ \bibinfo {author} {\bibfnamefont {A.}~\bibnamefont
  {Gali}},\ }\bibfield  {title} {\bibinfo {title} {Ab initio determination of
  pseudospin for paramagnetic defects in {S}i{C}},\ }\href
  {https://doi.org/10.1103/physrevb.102.241201} {\bibfield  {journal} {\bibinfo
   {journal} {Phys. Rev. B}\ }\textbf {\bibinfo {volume} {102}},\ \bibinfo
  {pages} {241201} (\bibinfo {year} {2020})}\BibitemShut {NoStop}%
\bibitem [{\citenamefont {Tissot}\ and\ \citenamefont
  {Burkard}(2021{\natexlab{a}})}]{tissot21a}%
  \BibitemOpen
  \bibfield  {author} {\bibinfo {author} {\bibfnamefont {B.}~\bibnamefont
  {Tissot}}\ and\ \bibinfo {author} {\bibfnamefont {G.}~\bibnamefont
  {Burkard}},\ }\bibfield  {title} {\bibinfo {title} {Spin structure and
  resonant driving of spin-{1/2} defects in {SiC}},\ }\href
  {https://doi.org/10.1103/physrevb.103.064106} {\bibfield  {journal} {\bibinfo
   {journal} {Phys. Rev. B}\ }\textbf {\bibinfo {volume} {103}},\ \bibinfo
  {pages} {064106} (\bibinfo {year} {2021}{\natexlab{a}})}\BibitemShut
  {NoStop}%
\bibitem [{\citenamefont {Tissot}\ and\ \citenamefont
  {Burkard}(2021{\natexlab{b}})}]{tissot21b}%
  \BibitemOpen
  \bibfield  {author} {\bibinfo {author} {\bibfnamefont {B.}~\bibnamefont
  {Tissot}}\ and\ \bibinfo {author} {\bibfnamefont {G.}~\bibnamefont
  {Burkard}},\ }\bibfield  {title} {\bibinfo {title} {Hyperfine structure of
  transition metal defects in {SiC}},\ }\href
  {https://doi.org/10.1103/physrevb.104.064102} {\bibfield  {journal} {\bibinfo
   {journal} {Phys. Rev. B}\ }\textbf {\bibinfo {volume} {104}},\ \bibinfo
  {pages} {064102} (\bibinfo {year} {2021}{\natexlab{b}})}\BibitemShut
  {NoStop}%
\bibitem [{\citenamefont {Gilardoni}\ \emph {et~al.}(2021)\citenamefont
  {Gilardoni}, \citenamefont {Ion}, \citenamefont {Hendriks}, \citenamefont
  {Trupke},\ and\ \citenamefont {van~der Wal}}]{gilardoni21}%
  \BibitemOpen
  \bibfield  {author} {\bibinfo {author} {\bibfnamefont {C.~M.}\ \bibnamefont
  {Gilardoni}}, \bibinfo {author} {\bibfnamefont {I.}~\bibnamefont {Ion}},
  \bibinfo {author} {\bibfnamefont {F.}~\bibnamefont {Hendriks}}, \bibinfo
  {author} {\bibfnamefont {M.}~\bibnamefont {Trupke}},\ and\ \bibinfo {author}
  {\bibfnamefont {C.~H.}\ \bibnamefont {van~der Wal}},\ }\bibfield  {title}
  {\bibinfo {title} {Hyperfine-mediated transitions between electronic spin-1/2
  levels of transition metal defects in {SiC}},\ }\bibfield  {journal}
  {\bibinfo  {journal} {New J. Phys.}\ }\href
  {https://doi.org/10.1088/1367-2630/ac1641} {10.1088/1367-2630/ac1641}
  (\bibinfo {year} {2021})\BibitemShut {NoStop}%
\bibitem [{\citenamefont {Tissot}\ \emph {et~al.}(2022)\citenamefont {Tissot},
  \citenamefont {Trupke}, \citenamefont {Koller}, \citenamefont {Astner},\ and\
  \citenamefont {Burkard}}]{tissot22}%
  \BibitemOpen
  \bibfield  {author} {\bibinfo {author} {\bibfnamefont {B.}~\bibnamefont
  {Tissot}}, \bibinfo {author} {\bibfnamefont {M.}~\bibnamefont {Trupke}},
  \bibinfo {author} {\bibfnamefont {P.}~\bibnamefont {Koller}}, \bibinfo
  {author} {\bibfnamefont {T.}~\bibnamefont {Astner}},\ and\ \bibinfo {author}
  {\bibfnamefont {G.}~\bibnamefont {Burkard}},\ }\bibfield  {title} {\bibinfo
  {title} {Nuclear spin quantum memory in silicon carbide},\ }\href
  {https://doi.org/10.1103/physrevresearch.4.033107} {\bibfield  {journal}
  {\bibinfo  {journal} {Phys. Rev. Res.}\ }\textbf {\bibinfo {volume} {4}},\
  \bibinfo {pages} {033107} (\bibinfo {year} {2022})}\BibitemShut {NoStop}%
\bibitem [{\citenamefont {Udvarhelyi}\ \emph {et~al.}(2018)\citenamefont
  {Udvarhelyi}, \citenamefont {Shkolnikov}, \citenamefont {Gali}, \citenamefont
  {Burkard},\ and\ \citenamefont {P{\'a}lyi}}]{udvarhelyi18}%
  \BibitemOpen
  \bibfield  {author} {\bibinfo {author} {\bibfnamefont {P.}~\bibnamefont
  {Udvarhelyi}}, \bibinfo {author} {\bibfnamefont {V.~O.}\ \bibnamefont
  {Shkolnikov}}, \bibinfo {author} {\bibfnamefont {A.}~\bibnamefont {Gali}},
  \bibinfo {author} {\bibfnamefont {G.}~\bibnamefont {Burkard}},\ and\ \bibinfo
  {author} {\bibfnamefont {A.}~\bibnamefont {P{\'a}lyi}},\ }\bibfield  {title}
  {\bibinfo {title} {Spin-strain interaction in nitrogen-vacancy centers in
  diamond},\ }\href {https://doi.org/10.1103/physrevb.98.075201} {\bibfield
  {journal} {\bibinfo  {journal} {Phys. Rev. B}\ }\textbf {\bibinfo {volume}
  {98}},\ \bibinfo {pages} {075201} (\bibinfo {year} {2018})}\BibitemShut
  {NoStop}%
\bibitem [{\citenamefont {Udvarhelyi}\ and\ \citenamefont
  {Gali}(2018)}]{Udvarhelyi_2018}%
  \BibitemOpen
  \bibfield  {author} {\bibinfo {author} {\bibfnamefont {P.}~\bibnamefont
  {Udvarhelyi}}\ and\ \bibinfo {author} {\bibfnamefont {A.}~\bibnamefont
  {Gali}},\ }\bibfield  {title} {\bibinfo {title} {Ab initio spin-strain
  coupling parameters of divacancy qubits in silicon carbide},\ }\href
  {https://doi.org/10.1103/PhysRevApplied.10.054010} {\bibfield  {journal}
  {\bibinfo  {journal} {Phys. Rev. Appl.}\ }\textbf {\bibinfo {volume} {10}},\
  \bibinfo {pages} {054010} (\bibinfo {year} {2018})}\BibitemShut {NoStop}%
\bibitem [{\citenamefont {Udvarhelyi}\ \emph {et~al.}(2020)\citenamefont
  {Udvarhelyi}, \citenamefont {Thiering}, \citenamefont {Morioka},
  \citenamefont {Babin}, \citenamefont {Kaiser}, \citenamefont {Lukin},
  \citenamefont {Ohshima}, \citenamefont {Ul-Hassan}, \citenamefont {Son},
  \citenamefont {Vu\ifmmode \check{c}\else
  \v{c}\fi{}kovi\ifmmode~\acute{c}\else \'{c}\fi{}}, \citenamefont
  {Wrachtrup},\ and\ \citenamefont {Gali}}]{Udvarhelyi_2020}%
  \BibitemOpen
  \bibfield  {author} {\bibinfo {author} {\bibfnamefont {P.}~\bibnamefont
  {Udvarhelyi}}, \bibinfo {author} {\bibfnamefont {G.}~\bibnamefont
  {Thiering}}, \bibinfo {author} {\bibfnamefont {N.}~\bibnamefont {Morioka}},
  \bibinfo {author} {\bibfnamefont {C.}~\bibnamefont {Babin}}, \bibinfo
  {author} {\bibfnamefont {F.}~\bibnamefont {Kaiser}}, \bibinfo {author}
  {\bibfnamefont {D.}~\bibnamefont {Lukin}}, \bibinfo {author} {\bibfnamefont
  {T.}~\bibnamefont {Ohshima}}, \bibinfo {author} {\bibfnamefont
  {J.}~\bibnamefont {Ul-Hassan}}, \bibinfo {author} {\bibfnamefont {N.~T.}\
  \bibnamefont {Son}}, \bibinfo {author} {\bibfnamefont {J.}~\bibnamefont
  {Vu\ifmmode \check{c}\else \v{c}\fi{}kovi\ifmmode~\acute{c}\else
  \'{c}\fi{}}}, \bibinfo {author} {\bibfnamefont {J.}~\bibnamefont
  {Wrachtrup}},\ and\ \bibinfo {author} {\bibfnamefont {A.}~\bibnamefont
  {Gali}},\ }\bibfield  {title} {\bibinfo {title} {Vibronic states and their
  effect on the temperature and strain dependence of silicon-vacancy qubits in
  4{H}-{S}i{C}},\ }\href {https://doi.org/10.1103/PhysRevApplied.13.054017}
  {\bibfield  {journal} {\bibinfo  {journal} {Phys. Rev. Appl.}\ }\textbf
  {\bibinfo {volume} {13}},\ \bibinfo {pages} {054017} (\bibinfo {year}
  {2020})}\BibitemShut {NoStop}%
\bibitem [{\citenamefont {Udvarhelyi}\ \emph {et~al.}(2023)\citenamefont
  {Udvarhelyi}, \citenamefont {Clua-Provost}, \citenamefont {Durand},
  \citenamefont {Li}, \citenamefont {Edgar}, \citenamefont {Gil}, \citenamefont
  {Cassabois}, \citenamefont {Jacques},\ and\ \citenamefont
  {Gali}}]{Udvarhelyi_2023}%
  \BibitemOpen
  \bibfield  {author} {\bibinfo {author} {\bibfnamefont {P.}~\bibnamefont
  {Udvarhelyi}}, \bibinfo {author} {\bibfnamefont {T.}~\bibnamefont
  {Clua-Provost}}, \bibinfo {author} {\bibfnamefont {A.}~\bibnamefont
  {Durand}}, \bibinfo {author} {\bibfnamefont {J.}~\bibnamefont {Li}}, \bibinfo
  {author} {\bibfnamefont {J.~H.}\ \bibnamefont {Edgar}}, \bibinfo {author}
  {\bibfnamefont {B.}~\bibnamefont {Gil}}, \bibinfo {author} {\bibfnamefont
  {G.}~\bibnamefont {Cassabois}}, \bibinfo {author} {\bibfnamefont
  {V.}~\bibnamefont {Jacques}},\ and\ \bibinfo {author} {\bibfnamefont
  {A.}~\bibnamefont {Gali}},\ }\href@noop {} {\bibinfo {title} {A planar defect
  spin sensor in a two-dimensional material susceptible to strain and electric
  fields}} (\bibinfo {year} {2023}),\ \Eprint
  {https://arxiv.org/abs/2304.00492} {arXiv:2304.00492 [quant-ph]} \BibitemShut
  {NoStop}%
\bibitem [{\citenamefont {Koppenh{\"o}fer}\ \emph {et~al.}(2023)\citenamefont
  {Koppenh{\"o}fer}, \citenamefont {Padgett}, \citenamefont {Cady},
  \citenamefont {Dharod}, \citenamefont {Oh}, \citenamefont {Jayich},\ and\
  \citenamefont {Clerk}}]{koppenhoefer23}%
  \BibitemOpen
  \bibfield  {author} {\bibinfo {author} {\bibfnamefont {M.}~\bibnamefont
  {Koppenh{\"o}fer}}, \bibinfo {author} {\bibfnamefont {C.}~\bibnamefont
  {Padgett}}, \bibinfo {author} {\bibfnamefont {J.~V.}\ \bibnamefont {Cady}},
  \bibinfo {author} {\bibfnamefont {V.}~\bibnamefont {Dharod}}, \bibinfo
  {author} {\bibfnamefont {H.}~\bibnamefont {Oh}}, \bibinfo {author}
  {\bibfnamefont {A.~C.~B.}\ \bibnamefont {Jayich}},\ and\ \bibinfo {author}
  {\bibfnamefont {A.~A.}\ \bibnamefont {Clerk}},\ }\bibfield  {title} {\bibinfo
  {title} {Single-spin readout and quantum sensing using optomechanically
  induced transparency},\ }\href
  {https://doi.org/10.1103/physrevlett.130.093603} {\bibfield  {journal}
  {\bibinfo  {journal} {Phys. Rev. Lett.}\ }\textbf {\bibinfo {volume} {130}},\
  \bibinfo {pages} {093603} (\bibinfo {year} {2023})}\BibitemShut {NoStop}%
\bibitem [{\citenamefont {Meesala}\ \emph {et~al.}(2018)\citenamefont
  {Meesala}, \citenamefont {Sohn}, \citenamefont {Pingault}, \citenamefont
  {Shao}, \citenamefont {Atikian}, \citenamefont {Holzgrafe}, \citenamefont
  {G{\"u}ndo{\u{g}}an}, \citenamefont {Stavrakas}, \citenamefont {Sipahigil},
  \citenamefont {Chia}, \citenamefont {Evans}, \citenamefont {Burek},
  \citenamefont {Zhang}, \citenamefont {Wu}, \citenamefont {Pacheco},
  \citenamefont {Abraham}, \citenamefont {Bielejec}, \citenamefont {Lukin},
  \citenamefont {Atat{\"u}re},\ and\ \citenamefont {Lon{\v{c}}ar}}]{meesala18}%
  \BibitemOpen
  \bibfield  {author} {\bibinfo {author} {\bibfnamefont {S.}~\bibnamefont
  {Meesala}}, \bibinfo {author} {\bibfnamefont {Y.-I.}\ \bibnamefont {Sohn}},
  \bibinfo {author} {\bibfnamefont {B.}~\bibnamefont {Pingault}}, \bibinfo
  {author} {\bibfnamefont {L.}~\bibnamefont {Shao}}, \bibinfo {author}
  {\bibfnamefont {H.~A.}\ \bibnamefont {Atikian}}, \bibinfo {author}
  {\bibfnamefont {J.}~\bibnamefont {Holzgrafe}}, \bibinfo {author}
  {\bibfnamefont {M.}~\bibnamefont {G{\"u}ndo{\u{g}}an}}, \bibinfo {author}
  {\bibfnamefont {C.}~\bibnamefont {Stavrakas}}, \bibinfo {author}
  {\bibfnamefont {A.}~\bibnamefont {Sipahigil}}, \bibinfo {author}
  {\bibfnamefont {C.}~\bibnamefont {Chia}}, \bibinfo {author} {\bibfnamefont
  {R.}~\bibnamefont {Evans}}, \bibinfo {author} {\bibfnamefont {M.~J.}\
  \bibnamefont {Burek}}, \bibinfo {author} {\bibfnamefont {M.}~\bibnamefont
  {Zhang}}, \bibinfo {author} {\bibfnamefont {L.}~\bibnamefont {Wu}}, \bibinfo
  {author} {\bibfnamefont {J.~L.}\ \bibnamefont {Pacheco}}, \bibinfo {author}
  {\bibfnamefont {J.}~\bibnamefont {Abraham}}, \bibinfo {author} {\bibfnamefont
  {E.}~\bibnamefont {Bielejec}}, \bibinfo {author} {\bibfnamefont {M.~D.}\
  \bibnamefont {Lukin}}, \bibinfo {author} {\bibfnamefont {M.}~\bibnamefont
  {Atat{\"u}re}},\ and\ \bibinfo {author} {\bibfnamefont {M.}~\bibnamefont
  {Lon{\v{c}}ar}},\ }\bibfield  {title} {\bibinfo {title} {Strain engineering
  of the silicon-vacancy center in diamond},\ }\href
  {https://doi.org/10.1103/physrevb.97.205444} {\bibfield  {journal} {\bibinfo
  {journal} {Physical Review B}\ }\textbf {\bibinfo {volume} {97}},\ \bibinfo
  {pages} {205444} (\bibinfo {year} {2018})}\BibitemShut {NoStop}%
\bibitem [{\citenamefont {Nguyen}\ \emph {et~al.}(2019)\citenamefont {Nguyen},
  \citenamefont {Sukachev}, \citenamefont {Bhaskar}, \citenamefont {Machielse},
  \citenamefont {Levonian}, \citenamefont {Knall}, \citenamefont {Stroganov},
  \citenamefont {Chia}, \citenamefont {Burek}, \citenamefont {Riedinger},
  \citenamefont {Park}, \citenamefont {Lon{\v{c}}ar},\ and\ \citenamefont
  {Lukin}}]{nguyen19}%
  \BibitemOpen
  \bibfield  {author} {\bibinfo {author} {\bibfnamefont {C.~T.}\ \bibnamefont
  {Nguyen}}, \bibinfo {author} {\bibfnamefont {D.~D.}\ \bibnamefont
  {Sukachev}}, \bibinfo {author} {\bibfnamefont {M.~K.}\ \bibnamefont
  {Bhaskar}}, \bibinfo {author} {\bibfnamefont {B.}~\bibnamefont {Machielse}},
  \bibinfo {author} {\bibfnamefont {D.~S.}\ \bibnamefont {Levonian}}, \bibinfo
  {author} {\bibfnamefont {E.~N.}\ \bibnamefont {Knall}}, \bibinfo {author}
  {\bibfnamefont {P.}~\bibnamefont {Stroganov}}, \bibinfo {author}
  {\bibfnamefont {C.}~\bibnamefont {Chia}}, \bibinfo {author} {\bibfnamefont
  {M.~J.}\ \bibnamefont {Burek}}, \bibinfo {author} {\bibfnamefont
  {R.}~\bibnamefont {Riedinger}}, \bibinfo {author} {\bibfnamefont
  {H.}~\bibnamefont {Park}}, \bibinfo {author} {\bibfnamefont {M.}~\bibnamefont
  {Lon{\v{c}}ar}},\ and\ \bibinfo {author} {\bibfnamefont {M.~D.}\ \bibnamefont
  {Lukin}},\ }\bibfield  {title} {\bibinfo {title} {An integrated nanophotonic
  quantum register based on silicon-vacancy spins in diamond},\ }\href
  {https://doi.org/10.1103/physrevb.100.165428} {\bibfield  {journal} {\bibinfo
   {journal} {Physical Review B}\ }\textbf {\bibinfo {volume} {100}},\ \bibinfo
  {pages} {165428} (\bibinfo {year} {2019})}\BibitemShut {NoStop}%
\bibitem [{\citenamefont {Ovartchaiyapong}\ \emph {et~al.}(2014)\citenamefont
  {Ovartchaiyapong}, \citenamefont {Lee}, \citenamefont {Myers},\ and\
  \citenamefont {Jayich}}]{ovartchaiyapong14}%
  \BibitemOpen
  \bibfield  {author} {\bibinfo {author} {\bibfnamefont {P.}~\bibnamefont
  {Ovartchaiyapong}}, \bibinfo {author} {\bibfnamefont {K.~W.}\ \bibnamefont
  {Lee}}, \bibinfo {author} {\bibfnamefont {B.~A.}\ \bibnamefont {Myers}},\
  and\ \bibinfo {author} {\bibfnamefont {A.~C.~B.}\ \bibnamefont {Jayich}},\
  }\bibfield  {title} {\bibinfo {title} {Dynamic strain-mediated coupling of a
  single diamond spin to a mechanical resonator},\ }\href
  {https://doi.org/10.1038/ncomms5429} {\bibfield  {journal} {\bibinfo
  {journal} {Nature Communications}\ }\textbf {\bibinfo {volume} {5}},\
  \bibinfo {pages} {4429} (\bibinfo {year} {2014})}\BibitemShut {NoStop}%
\bibitem [{\citenamefont {Barfuss}\ \emph {et~al.}(2019)\citenamefont
  {Barfuss}, \citenamefont {Kasperczyk}, \citenamefont {K{\"o}lbl},\ and\
  \citenamefont {Maletinsky}}]{barfuss19}%
  \BibitemOpen
  \bibfield  {author} {\bibinfo {author} {\bibfnamefont {A.}~\bibnamefont
  {Barfuss}}, \bibinfo {author} {\bibfnamefont {M.}~\bibnamefont {Kasperczyk}},
  \bibinfo {author} {\bibfnamefont {J.}~\bibnamefont {K{\"o}lbl}},\ and\
  \bibinfo {author} {\bibfnamefont {P.}~\bibnamefont {Maletinsky}},\ }\bibfield
   {title} {\bibinfo {title} {Spin-stress and spin-strain coupling in
  diamond-based hybrid spin oscillator systems},\ }\href
  {https://doi.org/10.1103/physrevb.99.174102} {\bibfield  {journal} {\bibinfo
  {journal} {Physical Review B}\ }\textbf {\bibinfo {volume} {99}},\ \bibinfo
  {pages} {174102} (\bibinfo {year} {2019})}\BibitemShut {NoStop}%
\bibitem [{Note1()}]{Note1}%
  \BibitemOpen
  \bibinfo {note} {According to previous DFT results the $k$ site corresponds
  to the \(\alpha \) configuration of vanadium in 4H-SiC \cite
  {csore20}.}\BibitemShut {Stop}%
\bibitem [{\citenamefont {Robledo}\ \emph {et~al.}(2011)\citenamefont
  {Robledo}, \citenamefont {Childress}, \citenamefont {Bernien}, \citenamefont
  {Hensen}, \citenamefont {Alkemade},\ and\ \citenamefont
  {Hanson}}]{robledo11}%
  \BibitemOpen
  \bibfield  {author} {\bibinfo {author} {\bibfnamefont {L.}~\bibnamefont
  {Robledo}}, \bibinfo {author} {\bibfnamefont {L.}~\bibnamefont {Childress}},
  \bibinfo {author} {\bibfnamefont {H.}~\bibnamefont {Bernien}}, \bibinfo
  {author} {\bibfnamefont {B.}~\bibnamefont {Hensen}}, \bibinfo {author}
  {\bibfnamefont {P.~F.~A.}\ \bibnamefont {Alkemade}},\ and\ \bibinfo {author}
  {\bibfnamefont {R.}~\bibnamefont {Hanson}},\ }\bibfield  {title} {\bibinfo
  {title} {High-fidelity projective read-out of a solid-state spin quantum
  register},\ }\href {https://doi.org/10.1038/nature10401} {\bibfield
  {journal} {\bibinfo  {journal} {Nature}\ }\textbf {\bibinfo {volume} {477}},\
  \bibinfo {pages} {574} (\bibinfo {year} {2011})}\BibitemShut {NoStop}%
\bibitem [{\citenamefont {Delteil}\ \emph {et~al.}(2014)\citenamefont
  {Delteil}, \citenamefont {Gao}, \citenamefont {Fallahi}, \citenamefont
  {Miguel-Sanchez},\ and\ \citenamefont {Imamo{\u{g}}lu}}]{delteil14}%
  \BibitemOpen
  \bibfield  {author} {\bibinfo {author} {\bibfnamefont {A.}~\bibnamefont
  {Delteil}}, \bibinfo {author} {\bibfnamefont {W.}~\bibnamefont {Gao}},
  \bibinfo {author} {\bibfnamefont {P.}~\bibnamefont {Fallahi}}, \bibinfo
  {author} {\bibfnamefont {J.}~\bibnamefont {Miguel-Sanchez}},\ and\ \bibinfo
  {author} {\bibfnamefont {A.}~\bibnamefont {Imamo{\u{g}}lu}},\ }\bibfield
  {title} {\bibinfo {title} {Observation of quantum jumps of a single quantum
  dot spin using submicrosecond single-shot optical readout},\ }\href
  {https://doi.org/10.1103/physrevlett.112.116802} {\bibfield  {journal}
  {\bibinfo  {journal} {Phys. Rev. Lett.}\ }\textbf {\bibinfo {volume} {112}},\
  \bibinfo {pages} {116802} (\bibinfo {year} {2014})}\BibitemShut {NoStop}%
\bibitem [{\citenamefont {Sukachev}\ \emph {et~al.}(2017)\citenamefont
  {Sukachev}, \citenamefont {Sipahigil}, \citenamefont {Nguyen}, \citenamefont
  {Bhaskar}, \citenamefont {Evans}, \citenamefont {Jelezko},\ and\
  \citenamefont {Lukin}}]{sukachev17}%
  \BibitemOpen
  \bibfield  {author} {\bibinfo {author} {\bibfnamefont {D.~D.}\ \bibnamefont
  {Sukachev}}, \bibinfo {author} {\bibfnamefont {A.}~\bibnamefont {Sipahigil}},
  \bibinfo {author} {\bibfnamefont {C.~T.}\ \bibnamefont {Nguyen}}, \bibinfo
  {author} {\bibfnamefont {M.~K.}\ \bibnamefont {Bhaskar}}, \bibinfo {author}
  {\bibfnamefont {R.~E.}\ \bibnamefont {Evans}}, \bibinfo {author}
  {\bibfnamefont {F.}~\bibnamefont {Jelezko}},\ and\ \bibinfo {author}
  {\bibfnamefont {M.~D.}\ \bibnamefont {Lukin}},\ }\bibfield  {title} {\bibinfo
  {title} {Silicon-vacancy spin qubit in diamond: a quantum memory exceeding 10
  ms with single-shot state readout},\ }\href
  {https://doi.org/10.1103/physrevlett.119.223602} {\bibfield  {journal}
  {\bibinfo  {journal} {Phys. Rev. Lett.}\ }\textbf {\bibinfo {volume} {119}},\
  \bibinfo {pages} {223602} (\bibinfo {year} {2017})}\BibitemShut {NoStop}%
\bibitem [{\citenamefont {Raha}\ \emph {et~al.}(2020)\citenamefont {Raha},
  \citenamefont {Chen}, \citenamefont {Phenicie}, \citenamefont {Ourari},
  \citenamefont {Dibos},\ and\ \citenamefont {Thompson}}]{raha20}%
  \BibitemOpen
  \bibfield  {author} {\bibinfo {author} {\bibfnamefont {M.}~\bibnamefont
  {Raha}}, \bibinfo {author} {\bibfnamefont {S.}~\bibnamefont {Chen}}, \bibinfo
  {author} {\bibfnamefont {C.~M.}\ \bibnamefont {Phenicie}}, \bibinfo {author}
  {\bibfnamefont {S.}~\bibnamefont {Ourari}}, \bibinfo {author} {\bibfnamefont
  {A.~M.}\ \bibnamefont {Dibos}},\ and\ \bibinfo {author} {\bibfnamefont
  {J.~D.}\ \bibnamefont {Thompson}},\ }\bibfield  {title} {\bibinfo {title}
  {Optical quantum nondemolition measurement of a single rare earth ion
  qubit},\ }\href {https://doi.org/10.1038/s41467-020-15138-7} {\bibfield
  {journal} {\bibinfo  {journal} {Nat. Commun.}\ }\textbf {\bibinfo {volume}
  {11}},\ \bibinfo {pages} {1605} (\bibinfo {year} {2020})}\BibitemShut
  {NoStop}%
\bibitem [{\citenamefont {Appel}\ \emph {et~al.}(2021)\citenamefont {Appel},
  \citenamefont {Tiranov}, \citenamefont {Javadi}, \citenamefont {L{\"o}bl},
  \citenamefont {Wang}, \citenamefont {Scholz}, \citenamefont {Wieck},
  \citenamefont {Ludwig}, \citenamefont {Warburton},\ and\ \citenamefont
  {Lodahl}}]{appel21}%
  \BibitemOpen
  \bibfield  {author} {\bibinfo {author} {\bibfnamefont {M.~H.}\ \bibnamefont
  {Appel}}, \bibinfo {author} {\bibfnamefont {A.}~\bibnamefont {Tiranov}},
  \bibinfo {author} {\bibfnamefont {A.}~\bibnamefont {Javadi}}, \bibinfo
  {author} {\bibfnamefont {M.~C.}\ \bibnamefont {L{\"o}bl}}, \bibinfo {author}
  {\bibfnamefont {Y.}~\bibnamefont {Wang}}, \bibinfo {author} {\bibfnamefont
  {S.}~\bibnamefont {Scholz}}, \bibinfo {author} {\bibfnamefont {A.~D.}\
  \bibnamefont {Wieck}}, \bibinfo {author} {\bibfnamefont {A.}~\bibnamefont
  {Ludwig}}, \bibinfo {author} {\bibfnamefont {R.~J.}\ \bibnamefont
  {Warburton}},\ and\ \bibinfo {author} {\bibfnamefont {P.}~\bibnamefont
  {Lodahl}},\ }\bibfield  {title} {\bibinfo {title} {Coherent spin-photon
  interface with waveguide induced cycling transitions},\ }\href
  {https://doi.org/10.1103/physrevlett.126.013602} {\bibfield  {journal}
  {\bibinfo  {journal} {Phys. Rev. Lett.}\ }\textbf {\bibinfo {volume} {126}},\
  \bibinfo {pages} {013602} (\bibinfo {year} {2021})}\BibitemShut {NoStop}%
\bibitem [{\citenamefont {Gray}\ \emph {et~al.}(1978)\citenamefont {Gray},
  \citenamefont {Whitley},\ and\ \citenamefont {Stroud}}]{gray78}%
  \BibitemOpen
  \bibfield  {author} {\bibinfo {author} {\bibfnamefont {H.~R.}\ \bibnamefont
  {Gray}}, \bibinfo {author} {\bibfnamefont {R.~M.}\ \bibnamefont {Whitley}},\
  and\ \bibinfo {author} {\bibfnamefont {C.~R.}\ \bibnamefont {Stroud}},\
  }\bibfield  {title} {\bibinfo {title} {Coherent trapping of atomic
  populations},\ }\href {https://doi.org/10.1364/ol.3.000218} {\bibfield
  {journal} {\bibinfo  {journal} {Opt. Lett.}\ }\textbf {\bibinfo {volume}
  {3}},\ \bibinfo {pages} {218} (\bibinfo {year} {1978})}\BibitemShut {NoStop}%
\bibitem [{\citenamefont {Xu}\ \emph {et~al.}(2008)\citenamefont {Xu},
  \citenamefont {Sun}, \citenamefont {Berman}, \citenamefont {Steel},
  \citenamefont {Bracker}, \citenamefont {Gammon},\ and\ \citenamefont
  {Sham}}]{xu08}%
  \BibitemOpen
  \bibfield  {author} {\bibinfo {author} {\bibfnamefont {X.}~\bibnamefont
  {Xu}}, \bibinfo {author} {\bibfnamefont {B.}~\bibnamefont {Sun}}, \bibinfo
  {author} {\bibfnamefont {P.~R.}\ \bibnamefont {Berman}}, \bibinfo {author}
  {\bibfnamefont {D.~G.}\ \bibnamefont {Steel}}, \bibinfo {author}
  {\bibfnamefont {A.~S.}\ \bibnamefont {Bracker}}, \bibinfo {author}
  {\bibfnamefont {D.}~\bibnamefont {Gammon}},\ and\ \bibinfo {author}
  {\bibfnamefont {L.~J.}\ \bibnamefont {Sham}},\ }\bibfield  {title} {\bibinfo
  {title} {Coherent population trapping of an electron spin in a single
  negatively charged quantum dot},\ }\href {https://doi.org/10.1038/nphys1054}
  {\bibfield  {journal} {\bibinfo  {journal} {Nature Phys.}\ }\textbf {\bibinfo
  {volume} {4}},\ \bibinfo {pages} {692} (\bibinfo {year} {2008})}\BibitemShut
  {NoStop}%
\bibitem [{\citenamefont {Kelly}\ \emph {et~al.}(2010)\citenamefont {Kelly},
  \citenamefont {Dutton}, \citenamefont {Schlafer}, \citenamefont {Mookerji},
  \citenamefont {Ohki}, \citenamefont {Kline},\ and\ \citenamefont
  {Pappas}}]{kelly10}%
  \BibitemOpen
  \bibfield  {author} {\bibinfo {author} {\bibfnamefont {W.~R.}\ \bibnamefont
  {Kelly}}, \bibinfo {author} {\bibfnamefont {Z.}~\bibnamefont {Dutton}},
  \bibinfo {author} {\bibfnamefont {J.}~\bibnamefont {Schlafer}}, \bibinfo
  {author} {\bibfnamefont {B.}~\bibnamefont {Mookerji}}, \bibinfo {author}
  {\bibfnamefont {T.~A.}\ \bibnamefont {Ohki}}, \bibinfo {author}
  {\bibfnamefont {J.~S.}\ \bibnamefont {Kline}},\ and\ \bibinfo {author}
  {\bibfnamefont {D.~P.}\ \bibnamefont {Pappas}},\ }\bibfield  {title}
  {\bibinfo {title} {Direct observation of coherent population trapping in a
  superconducting artificial atom},\ }\href
  {https://doi.org/10.1103/physrevlett.104.163601} {\bibfield  {journal}
  {\bibinfo  {journal} {Phys. Rev. Lett.}\ }\textbf {\bibinfo {volume} {104}},\
  \bibinfo {pages} {163601} (\bibinfo {year} {2010})}\BibitemShut {NoStop}%
\bibitem [{\citenamefont {Dong}\ \emph {et~al.}(2012)\citenamefont {Dong},
  \citenamefont {Fiore}, \citenamefont {Kuzyk},\ and\ \citenamefont
  {Wang}}]{dong12}%
  \BibitemOpen
  \bibfield  {author} {\bibinfo {author} {\bibfnamefont {C.}~\bibnamefont
  {Dong}}, \bibinfo {author} {\bibfnamefont {V.}~\bibnamefont {Fiore}},
  \bibinfo {author} {\bibfnamefont {M.~C.}\ \bibnamefont {Kuzyk}},\ and\
  \bibinfo {author} {\bibfnamefont {H.}~\bibnamefont {Wang}},\ }\bibfield
  {title} {\bibinfo {title} {Optomechanical dark mode},\ }\href
  {https://doi.org/10.1126/science.1228370} {\bibfield  {journal} {\bibinfo
  {journal} {Science}\ }\textbf {\bibinfo {volume} {338}},\ \bibinfo {pages}
  {1609} (\bibinfo {year} {2012})}\BibitemShut {NoStop}%
\bibitem [{\citenamefont {Demirplak}\ and\ \citenamefont
  {Rice}(2003)}]{demirplak03}%
  \BibitemOpen
  \bibfield  {author} {\bibinfo {author} {\bibfnamefont {M.}~\bibnamefont
  {Demirplak}}\ and\ \bibinfo {author} {\bibfnamefont {S.~A.}\ \bibnamefont
  {Rice}},\ }\bibfield  {title} {\bibinfo {title} {Adiabatic population
  transfer with control fields},\ }\href {https://doi.org/10.1021/jp030708a}
  {\bibfield  {journal} {\bibinfo  {journal} {J. Phys. Chem. A}\ }\textbf
  {\bibinfo {volume} {107}},\ \bibinfo {pages} {9937} (\bibinfo {year}
  {2003})}\BibitemShut {NoStop}%
\bibitem [{\citenamefont {Berry}(2009)}]{berry09}%
  \BibitemOpen
  \bibfield  {author} {\bibinfo {author} {\bibfnamefont {M.~V.}\ \bibnamefont
  {Berry}},\ }\bibfield  {title} {\bibinfo {title} {Transitionless quantum
  driving},\ }\href {https://doi.org/10.1088/1751-8113/42/36/365303} {\bibfield
   {journal} {\bibinfo  {journal} {J. Phys. A: Math. Theor.}\ }\textbf
  {\bibinfo {volume} {42}},\ \bibinfo {pages} {365303} (\bibinfo {year}
  {2009})}\BibitemShut {NoStop}%
\bibitem [{\citenamefont {Gu{\'e}ry-Odelin}\ \emph {et~al.}(2019)\citenamefont
  {Gu{\'e}ry-Odelin}, \citenamefont {Ruschhaupt}, \citenamefont {Kiely},
  \citenamefont {Torrontegui}, \citenamefont {Mart{\'i}nez-Garaot},\ and\
  \citenamefont {Muga}}]{guery-odelin19}%
  \BibitemOpen
  \bibfield  {author} {\bibinfo {author} {\bibfnamefont {D.}~\bibnamefont
  {Gu{\'e}ry-Odelin}}, \bibinfo {author} {\bibfnamefont {A.}~\bibnamefont
  {Ruschhaupt}}, \bibinfo {author} {\bibfnamefont {A.}~\bibnamefont {Kiely}},
  \bibinfo {author} {\bibfnamefont {E.}~\bibnamefont {Torrontegui}}, \bibinfo
  {author} {\bibfnamefont {S.}~\bibnamefont {Mart{\'i}nez-Garaot}},\ and\
  \bibinfo {author} {\bibfnamefont {J.~G.}\ \bibnamefont {Muga}},\ }\bibfield
  {title} {\bibinfo {title} {Shortcuts to adiabaticity: Concepts, methods, and
  applications},\ }\href {https://doi.org/10.1103/revmodphys.91.045001}
  {\bibfield  {journal} {\bibinfo  {journal} {Rev. Mod. Phys.}\ }\textbf
  {\bibinfo {volume} {91}},\ \bibinfo {pages} {045001} (\bibinfo {year}
  {2019})}\BibitemShut {NoStop}%
\bibitem [{\citenamefont {Hendriks}\ \emph {et~al.}(2022)\citenamefont
  {Hendriks}, \citenamefont {Gilardoni}, \citenamefont {Adambukulam},
  \citenamefont {Laucht},\ and\ \citenamefont {Wal}}]{hendriks22}%
  \BibitemOpen
  \bibfield  {author} {\bibinfo {author} {\bibfnamefont {J.}~\bibnamefont
  {Hendriks}}, \bibinfo {author} {\bibfnamefont {C.~M.}\ \bibnamefont
  {Gilardoni}}, \bibinfo {author} {\bibfnamefont {C.}~\bibnamefont
  {Adambukulam}}, \bibinfo {author} {\bibfnamefont {A.}~\bibnamefont
  {Laucht}},\ and\ \bibinfo {author} {\bibfnamefont {C.~H. v.~d.}\ \bibnamefont
  {Wal}},\ }\bibfield  {title} {\bibinfo {title} {Coherent spin dynamics of
  hyperfine-coupled vanadium impurities in silicon carbide},\ }\href
  {http://arxiv.org/abs/2210.09942v1} {\bibfield  {journal} {\bibinfo
  {journal} {CoRR}\ } (\bibinfo {year} {2022})},\ \Eprint
  {https://arxiv.org/abs/2210.09942} {arXiv:2210.09942 [quant-ph]} \BibitemShut
  {NoStop}%
\bibitem [{\citenamefont {Kresse}\ and\ \citenamefont {Hafner}(1993)}]{VASP1}%
  \BibitemOpen
  \bibfield  {author} {\bibinfo {author} {\bibfnamefont {G.}~\bibnamefont
  {Kresse}}\ and\ \bibinfo {author} {\bibfnamefont {J.}~\bibnamefont
  {Hafner}},\ }\bibfield  {title} {\bibinfo {title} {\textit{Ab initio}
  molecular dynamics for liquid metals},\ }\href
  {https://doi.org/10.1103/PhysRevB.47.558} {\bibfield  {journal} {\bibinfo
  {journal} {Phys. Rev. B}\ }\textbf {\bibinfo {volume} {47}},\ \bibinfo
  {pages} {558} (\bibinfo {year} {1993})}\BibitemShut {NoStop}%
\bibitem [{\citenamefont {Kresse}\ and\ \citenamefont
  {Furthm\"uller}(1996{\natexlab{a}})}]{VASP2}%
  \BibitemOpen
  \bibfield  {author} {\bibinfo {author} {\bibfnamefont {G.}~\bibnamefont
  {Kresse}}\ and\ \bibinfo {author} {\bibfnamefont {J.}~\bibnamefont
  {Furthm\"uller}},\ }\bibfield  {title} {\bibinfo {title} {Efficient iterative
  schemes for \textit{ab initio} total-energy calculations using a plane-wave
  basis set},\ }\href {https://doi.org/10.1103/PhysRevB.54.11169} {\bibfield
  {journal} {\bibinfo  {journal} {Phys. Rev. B}\ }\textbf {\bibinfo {volume}
  {54}},\ \bibinfo {pages} {11169} (\bibinfo {year}
  {1996}{\natexlab{a}})}\BibitemShut {NoStop}%
\bibitem [{\citenamefont {Kresse}\ and\ \citenamefont
  {Furthm\"uller}(1996{\natexlab{b}})}]{VASP3}%
  \BibitemOpen
  \bibfield  {author} {\bibinfo {author} {\bibfnamefont {G.}~\bibnamefont
  {Kresse}}\ and\ \bibinfo {author} {\bibfnamefont {J.}~\bibnamefont
  {Furthm\"uller}},\ }\bibfield  {title} {\bibinfo {title} {Efficiency of
  ab-initio total energy calculations for metals and semiconductors using a
  plane-wave basis set},\ }\href
  {https://doi.org/http://dx.doi.org/10.1016/0927-0256(96)00008-0} {\bibfield
  {journal} {\bibinfo  {journal} {Comput. Mater. Sci.}\ }\textbf {\bibinfo
  {volume} {6}},\ \bibinfo {pages} {15 } (\bibinfo {year}
  {1996}{\natexlab{b}})}\BibitemShut {NoStop}%
\bibitem [{\citenamefont {Paier}\ \emph {et~al.}(2006)\citenamefont {Paier},
  \citenamefont {Marsman}, \citenamefont {Hummer}, \citenamefont {Kresse},
  \citenamefont {Gerber},\ and\ \citenamefont {\'Angy\'an}}]{VASP4}%
  \BibitemOpen
  \bibfield  {author} {\bibinfo {author} {\bibfnamefont {J.}~\bibnamefont
  {Paier}}, \bibinfo {author} {\bibfnamefont {M.}~\bibnamefont {Marsman}},
  \bibinfo {author} {\bibfnamefont {K.}~\bibnamefont {Hummer}}, \bibinfo
  {author} {\bibfnamefont {G.}~\bibnamefont {Kresse}}, \bibinfo {author}
  {\bibfnamefont {I.~C.}\ \bibnamefont {Gerber}},\ and\ \bibinfo {author}
  {\bibfnamefont {J.~G.}\ \bibnamefont {\'Angy\'an}},\ }\bibfield  {title}
  {\bibinfo {title} {Screened hybrid density functionals applied to solids},\
  }\href {https://doi.org/10.1063/1.2187006} {\bibfield  {journal} {\bibinfo
  {journal} {J. Chem. Phys.}\ }\textbf {\bibinfo {volume} {124}},\ \bibinfo
  {pages} {154709} (\bibinfo {year} {2006})}\BibitemShut {NoStop}%
\bibitem [{\citenamefont {Bl\"ochl}(1994)}]{PAW}%
  \BibitemOpen
  \bibfield  {author} {\bibinfo {author} {\bibfnamefont {P.~E.}\ \bibnamefont
  {Bl\"ochl}},\ }\bibfield  {title} {\bibinfo {title} {Projector augmented-wave
  method},\ }\href {https://doi.org/10.1103/PhysRevB.50.17953} {\bibfield
  {journal} {\bibinfo  {journal} {Phys. Rev. B}\ }\textbf {\bibinfo {volume}
  {50}},\ \bibinfo {pages} {17953} (\bibinfo {year} {1994})}\BibitemShut
  {NoStop}%
\bibitem [{\citenamefont {Krukau}\ \emph {et~al.}(2006)\citenamefont {Krukau},
  \citenamefont {Vydrov}, \citenamefont {Izmaylov},\ and\ \citenamefont
  {Scuseria}}]{HSE06}%
  \BibitemOpen
  \bibfield  {author} {\bibinfo {author} {\bibfnamefont {A.~V.}\ \bibnamefont
  {Krukau}}, \bibinfo {author} {\bibfnamefont {O.~A.}\ \bibnamefont {Vydrov}},
  \bibinfo {author} {\bibfnamefont {A.~F.}\ \bibnamefont {Izmaylov}},\ and\
  \bibinfo {author} {\bibfnamefont {G.~E.}\ \bibnamefont {Scuseria}},\
  }\bibfield  {title} {\bibinfo {title} {Influence of the exchange screening
  parameter on the performance of screened hybrid functionals},\ }\href
  {https://doi.org/10.1063/1.2404663} {\bibfield  {journal} {\bibinfo
  {journal} {J. Chem. Phys.}\ }\textbf {\bibinfo {volume} {125}},\ \bibinfo
  {pages} {224106} (\bibinfo {year} {2006})}\BibitemShut {NoStop}%
\bibitem [{\citenamefont {Dudarev}\ \emph {et~al.}(1998)\citenamefont
  {Dudarev}, \citenamefont {Botton}, \citenamefont {Savrasov}, \citenamefont
  {Humphreys},\ and\ \citenamefont {Sutton}}]{Dudarev_1998}%
  \BibitemOpen
  \bibfield  {author} {\bibinfo {author} {\bibfnamefont {S.~L.}\ \bibnamefont
  {Dudarev}}, \bibinfo {author} {\bibfnamefont {G.~A.}\ \bibnamefont {Botton}},
  \bibinfo {author} {\bibfnamefont {S.~Y.}\ \bibnamefont {Savrasov}}, \bibinfo
  {author} {\bibfnamefont {C.~J.}\ \bibnamefont {Humphreys}},\ and\ \bibinfo
  {author} {\bibfnamefont {A.~P.}\ \bibnamefont {Sutton}},\ }\bibfield  {title}
  {\bibinfo {title} {Electron-energy-loss spectra and the structural stability
  of nickel oxide: An {LSDA+U} study},\ }\href
  {https://doi.org/10.1103/PhysRevB.57.1505} {\bibfield  {journal} {\bibinfo
  {journal} {Phys. Rev. B}\ }\textbf {\bibinfo {volume} {57}},\ \bibinfo
  {pages} {1505} (\bibinfo {year} {1998})}\BibitemShut {NoStop}%
\bibitem [{\citenamefont {Iv\'ady}\ \emph {et~al.}(2013)\citenamefont
  {Iv\'ady}, \citenamefont {Abrikosov}, \citenamefont {Janz\'en},\ and\
  \citenamefont {Gali}}]{Ivady_2013}%
  \BibitemOpen
  \bibfield  {author} {\bibinfo {author} {\bibfnamefont {V.}~\bibnamefont
  {Iv\'ady}}, \bibinfo {author} {\bibfnamefont {I.~A.}\ \bibnamefont
  {Abrikosov}}, \bibinfo {author} {\bibfnamefont {E.}~\bibnamefont
  {Janz\'en}},\ and\ \bibinfo {author} {\bibfnamefont {A.}~\bibnamefont
  {Gali}},\ }\bibfield  {title} {\bibinfo {title} {Role of screening in the
  density functional applied to transition-metal defects in semiconductors},\
  }\href {https://doi.org/10.1103/PhysRevB.87.205201} {\bibfield  {journal}
  {\bibinfo  {journal} {Phys. Rev. B}\ }\textbf {\bibinfo {volume} {87}},\
  \bibinfo {pages} {205201} (\bibinfo {year} {2013})}\BibitemShut {NoStop}%
\bibitem [{\citenamefont {Gali}\ \emph {et~al.}(2009)\citenamefont {Gali},
  \citenamefont {Janz\'en}, \citenamefont {De\'ak}, \citenamefont {Kresse},\
  and\ \citenamefont {Kaxiras}}]{Gali_2009}%
  \BibitemOpen
  \bibfield  {author} {\bibinfo {author} {\bibfnamefont {A.}~\bibnamefont
  {Gali}}, \bibinfo {author} {\bibfnamefont {E.}~\bibnamefont {Janz\'en}},
  \bibinfo {author} {\bibfnamefont {P.}~\bibnamefont {De\'ak}}, \bibinfo
  {author} {\bibfnamefont {G.}~\bibnamefont {Kresse}},\ and\ \bibinfo {author}
  {\bibfnamefont {E.}~\bibnamefont {Kaxiras}},\ }\bibfield  {title} {\bibinfo
  {title} {Theory of spin-conserving excitation of the {N}{V}$^{-}$ center in
  diamond},\ }\href {https://doi.org/10.1103/PhysRevLett.103.186404} {\bibfield
   {journal} {\bibinfo  {journal} {Phys. Rev. Lett.}\ }\textbf {\bibinfo
  {volume} {103}},\ \bibinfo {pages} {186404} (\bibinfo {year}
  {2009})}\BibitemShut {NoStop}%
\bibitem [{\citenamefont {Bravyi}\ \emph {et~al.}(2011)\citenamefont {Bravyi},
  \citenamefont {DiVincenzo},\ and\ \citenamefont {Loss}}]{bravyi11}%
  \BibitemOpen
  \bibfield  {author} {\bibinfo {author} {\bibfnamefont {S.}~\bibnamefont
  {Bravyi}}, \bibinfo {author} {\bibfnamefont {D.~P.}\ \bibnamefont
  {DiVincenzo}},\ and\ \bibinfo {author} {\bibfnamefont {D.}~\bibnamefont
  {Loss}},\ }\bibfield  {title} {\bibinfo {title} {{Schrieffer-Wolff}
  transformation for quantum many-body systems},\ }\href
  {https://doi.org/10.1016/j.aop.2011.06.004} {\bibfield  {journal} {\bibinfo
  {journal} {Ann. Phys.}\ }\textbf {\bibinfo {volume} {326}},\ \bibinfo {pages}
  {2793} (\bibinfo {year} {2011})}\BibitemShut {NoStop}%
\end{thebibliography}%

\end{document}